\documentclass[12pt, draftclsnofoot, onecolumn]{IEEEtran}

\usepackage{booktabs} 

\usepackage{amsmath}
\usepackage{amssymb}
\usepackage{amsthm}
\usepackage{xfrac}
\usepackage{color}
\usepackage{graphicx}
\usepackage{stfloats}
\usepackage{algorithm}
\usepackage{algorithmic}
\usepackage{flushend}
\usepackage{caption}
\usepackage{enumitem}

\usepackage{amsmath}
\usepackage{amssymb}
\usepackage[normalem]{ulem}
\usepackage{graphicx}
\usepackage{flushend}
\usepackage{url}
\usepackage{color}
\usepackage{algorithm}
\usepackage{algorithmic}
\usepackage{ragged2e}
\usepackage{amssymb}
\usepackage{fancyvrb}

\usepackage[utf8]{inputenc}
\usepackage{pgfplots}
\usepackage{amsmath}

\usepackage{subfig}
\usetikzlibrary{patterns}

\usepackage[numbers,sort&compress]{natbib}

\DeclareMathOperator*{\argmax}{argmax}

\begin{document}
\pagestyle{plain} 

\graphicspath{{figure/}{figures/}}

\newcommand{\mf}{\mathbf}
\newcommand{\mc}{\mathcal}
\newcommand{\mb}{\mathbb}
\newcommand{\mr}{\mathrm}
\newcommand{\bs}{\boldsymbol}
\newcommand{\st}{\mathrm{subject\ to}}
\newcommand{\minimize}{\mathrm{minimize}}

\newcommand{\todo}[1]{\textbf{\color{red}[XX #1 XX]}}


\newcommand{\norm}[1]{\left\lVert#1\right\rVert}

\hyphenation{fin-ger-prin-ting fil-te-ring-based}

\title{\textit{Hacking the Waveform:} Generalized Wireless Adversarial Deep Learning}

\author{Francesco Restuccia, Salvatore D'Oro, Amani Al-Shawabka, Bruno Costa Rendon, Kaushik Chowdhury, Stratis Ioannidis, and  Tommaso Melodia\vspace{-1.6cm}  
\thanks{This paper has been submitted for possible publication to IEEE Transactions on Wireless Communications. The authors are with the Institute for the Wireless Internet of Things, Department of Electrical and Computer Engineering, Northeastern University, Boston, MA, 02215 USA. Corresponding author e-mail: melodia@northeastern.edu.}}%

\maketitle

\begin{abstract}
Deep learning techniques can classify spectrum phenomena (\textit{e.g.}, waveform modulation) with accuracy levels that were once thought impossible. Although we have recently seen many advances in this field, extensive work in computer vision has demonstrated that adversarial machine learning (AML) can seriously decrease the accuracy of a classifier. This is done by designing inputs that are close to a legitimate one but interpreted by the classifier as being of a completely different class. On the other hand, it is unclear \emph{if}, \emph{when}, and \emph{how} AML is concretely possible in practical wireless scenarios, where (i) the highly time-varying nature of the channel could compromise adversarial attempts; and (ii) the received waveforms still need to be decodable and thus cannot be extensively modified. 
This paper advances the state of the art by proposing the first comprehensive analysis and experimental evaluation of adversarial learning attacks to wireless deep learning systems. We postulate a series of adversarial attacks, and formulate a Generalized Wireless Adversarial Machine Learning Problem (GWAP) where we analyze the combined effect of the wireless channel and the adversarial waveform on the efficacy of the attacks. We propose a new neural network architecture called \emph{FIRNet}, which can be trained to ``hack'' a classifier based only on its output. We extensively evaluate the performance on (i) a 1,000-device radio fingerprinting dataset, and (ii) a 24-class modulation dataset. Results obtained with several channel conditions show that our algorithms can decrease the classifier accuracy up to 3x. We also experimentally evaluate \emph{FIRNet} on a radio testbed, and show that our data-driven blackbox approach can confuse the classifier up to 97\% while keeping the waveform distortion to a minimum. \vspace{-0.2cm}

\end{abstract}

\section{Introduction}

The Internet of Things (IoT) will bring 75.44B devices on the market by 2025, a 5x increase in ten years \cite{Statista-IoT}. Due to the sheer number of IoT devices soon to be deployed worldwide, the design of practical spectrum knowledge extraction techniques has now become a compelling necessity --  not only to understand in real time the wireless environment, but also to design reactive, intelligent, and more secure wireless protocols, systems, and architectures \cite{Restuccia-infocom2019}.

Arguably, the radio frequency (RF) spectrum is one of nature's most complex phenomena. For this reason, the wireless community has started to move toward data-driven solutions based on \textit{deep learning} \cite{lecun2015deep} -- well-known to be exceptionally suited to solve classification problems where a mathematical model is impossible to obtain. Extensively applied since the 1980s,  neural networks are now being used to address notoriously hard problems such as radio fingerprinting \cite{restuccia2019deepradioid}, signal/traffic classification \cite{OShea-ieeejstsp2018,Restuccia-infocom2019,restuccia2020polymorf} and resource allocation  \cite{Zhang-infocom2019}, among many others \cite{jagannath2019machine}.

Recent advances in wireless deep learning have now clearly demonstrated its great potential. For example, O'Shea \emph{et al.} \cite{OShea-ieeejstsp2018} has demonstrated that models based on deep learning can achieve about 20\% higher modulation classification accuracy than legacy learning models under noisy channel conditions. However, it has been extensively proven that neural networks are prone to be ``hacked'' by carefully crafting small-scale perturbations to the input -- which keep the input similar to the original one, but are ultimately able to ``steer'' the neural network away from the ground truth. This activity is  known  \cite{restuccia2020wiseml,Goodfellow-iclr2015,Moosavi-cvpr2017,Papernot-asiaccs2017,Carlini-ieeesp2017} as \textit{adversarial machine learning} (AML). The degree to which malicious agents can find adversarial examples is strongly correlated to the applicability of neural networks to address problems in the wireless domain \cite{Restuccia-IoT2018}. \smallskip

\textbf{Technical Challenges.~}We believe the above reasons clearly show the timeliness and urgency of a rigorous investigation into the \emph{robustness} of wireless deep learning systems. Prior work \cite{Shi-wiseml2019,Bair-wiseml2019}  -- which is discussed in great detail in Section \ref{sec:rw} -- is severely limited by small-scale simulation-based scenarios, which has left several fundamental questions unanswered. The key reason that sets wireless AML apart is that a wireless deep learning system is affected by the stochastic nature of the channel \cite{goldsmith2005wireless}. This implies that the channel action must be factored into the crafting process of the AML attack.  

\begin{figure}[!h]
    \centering
    \includegraphics[width=\columnwidth]{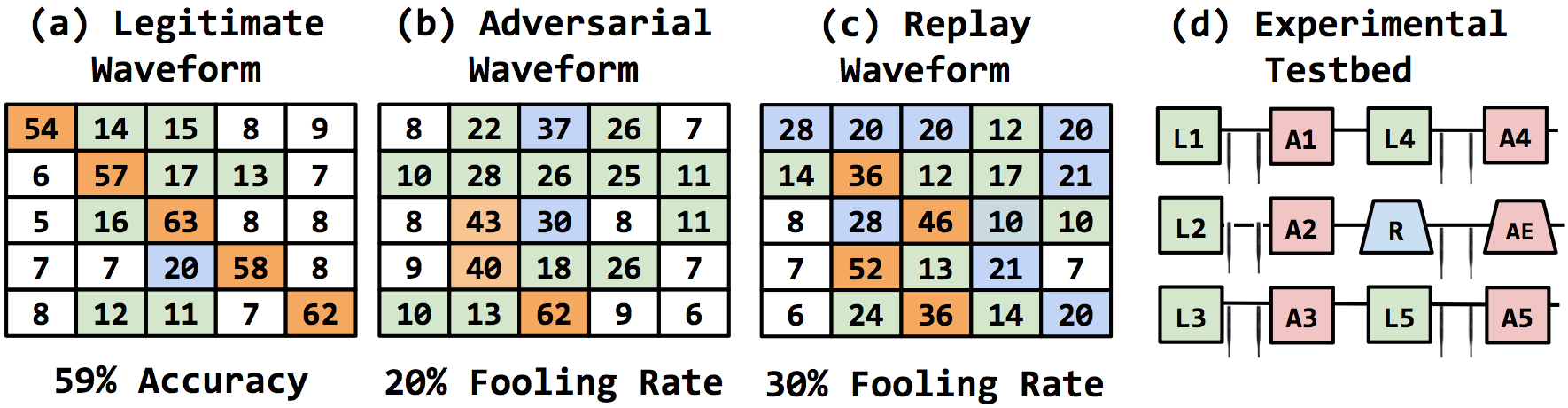}
    \caption{Simple attacks may not hack a wireless classifier.\vspace{-0.5cm}}
    \label{fig:conf_mat_intro}
\end{figure}

To further confirm this critical aspect, Figure \ref{fig:conf_mat_intro} reports a series of experiments results obtained with our software-defined radio testbed (see Section \ref{sec:over_the_air}). In our setup shown in Figure \ref{fig:conf_mat_intro}(d), we collect a series of waveforms coming from 5 legitimate transmitters (L1 to L5) through a legitimate receiver (R). Then, we train a neural network (see Section \ref{sec:over_the_air})  to recognize the legitimate devices by learning the unique impairments imposed by the radio circuitry on the transmitted waveforms, also called \textit{radio fingerprinting} \cite{restuccia2019deepradioid}. The neural network obtains 59\% accuracy, as shown in \ref{fig:conf_mat_intro}(a). We also use an adversarial eavesdropper radio (AE) to record the waveforms transmitted by the legitimate transmitters. We show the fooling rate obtained by 5 adversarial devices A1 to A5 which transmit RF waveforms trying to fool the classifier by imitating respectively L1 to L5. A high fooling rate means that adversaries can generate waveforms that are classified as belonging to legitimate devices. On the contrary, a low fooling rate indicates that the attack is unsuccessful as the classifier is not able to identify received waveforms. In this experiment, we consider two substantially different attacks where adversaries (i) transmit their own waveforms  -- shown in \ref{fig:conf_mat_intro}(b); and (ii) ``replay'' the recorded waveforms from L1 to L5 (i.e., by simply retransmitting the I/Q samples recorder by the eavesdropper AE) -- shown in \ref{fig:conf_mat_intro}(c). Figure \ref{fig:conf_mat_intro}(b) shows that when A1 to A5 transmit their own waveforms, the fooling rate is 20\%, way lower than the original accuracy of 59\%. In principle, we would expect the adversary to obtain a significant increase in fooling rate by performing the replay attack. However, \ref{fig:conf_mat_intro}(c) indicates that the fooling rate is only 30\% when A1 to A5 replay the eavesdropped waveforms. This strongly suggests that even if the adversary is successful in replaying the waveforms, the channel will inevitably make the attack less effective. Thus, more complex attacks have to be designed and tested to validate whether AML is effectively a threat in the wireless domain.



\smallskip 

\textbf{Novel Contributions.}~The key contribution of this paper is to provide the first comprehensive modeling and experimental evaluation of adversarial machine learning (AML) attacks to state-of-the-art wireless deep learning systems. To this end, our study bridges together concepts from both the wireless and the adversarial learning domains, which have been so far kept  separated. 

We summarize our core technical contributions as follows:\smallskip

$\bullet$ We propose a novel AML threat model (Section \ref{sec:threat_model}) where we consider (i) a ``whitebox'' scenario, where the adversary has complete access to the neural network; and (ii) a ``blackbox'' scenario, where the neural network is not available to the adversary. The primary advance of our model is that our attacks are derived for arbitrary channels, waveforms, and neural networks, and thus generalizable to any state-of-the-art wireless deep learning system;\smallskip

$\bullet$ Based on the proposed model, we formulate an \textit{AML Waveform Jamming} (Section \ref{sec:jamming}) and an \textit{AML Waveform Synthesis} (Section \ref{sec:synt}) attack. Next, we propose a \textit{Generalized Wireless Adversarial Machine Learning Problem} (GWAP) where an adversary tries to steer the neural network away from the ground truth while satisfying constraints such as bit error rate, radiated power, and other relevant metrics below a threshold (Section \ref{sec:optimization}). Next, we propose in Section \ref{sec:gradient_based} a gradient-based algorithm to solve the GWAP in a whitebox scenario. For the blackbox scenario, we design a novel neural network architecture called \textit{FIRNet}. Our approach mixes together concepts from generative adversarial learning and signal processing to train a neural network composed by finite impulse response layers (FIRLayers), which are trained to impose small-scale modifications to the input and at the same time decrease the classifier's accuracy; \smallskip

$\bullet$ We extensively evaluate the proposed algorithms on (i) a deep learning model for radio fingerprinting \cite{restuccia2019deepradioid} trained on a 1,000-device dataset of WiFi and ADS-B transmissions collected \textit{in the wild}; and (ii) a modulation recognition model \cite{OShea-ieeejstsp2018} trained on the widely-available RadioML 2018.01A dataset, which includes 24 different analog and digital modulations with different levels of signal-to-noise ratio (SNR). Our algorithms are shown to decrease the accuracy of the models up to 3x in case of whitebox attacks, while keeping the waveform distortion to a minimum. Moreover, we evaluate our \textit{FIRNet} approach on the software-defined radio testbed, and show that our approach confuses the 5-device radio fingerprinting classifier up to 97\%.

\section{Related Work}\label{sec:rw}



Adversarial machine learning (AML) has been extensively investigated in computer vision.  Szegedy \textit{et al.} \cite{Szegedy-2013intriguing} first pointed out the existence of \textit{targeted} adversarial examples: given a valid input $x$, a classifier $C$ and a target $t$, it is possible to find $x' \sim x$ such that $C(x') = t$. More recently, Moosavi-Dezfooli \emph{et al.} \cite{Moosavi-cvpr2017} have further demonstrated the existence of so-called \textit{universal perturbation vectors}, such that for the majority of inputs $x$, it holds that $C(x + v) \not= C(x)$. Carlini and Kruger \cite{Carlini-ieeesp2017} evaluated a series of adversarial attacks that are shown to be effective against defensive neural network distillation \cite{Papernot-ieeesp2016}. Although the above papers have made significant advances in our understanding of AML, it can only be applied to stationary learning contexts such as computer vision. The presence of non-stationarity makes wireless AML significantly more challenging and thus worth of additional investigation.

Only very recently has AML been approached by the wireless community. Bair \emph{et al.} \cite{Bair-wiseml2019} propose to apply a variation of the MI-FGSM attack \cite{dong2018boosting} to create adversarial examples to modulation classification systems. Shi \emph{et al.} \cite{Shi-wiseml2019} propose the usage of a generative adversarial network (GAN) to spoof a targeted device. However, the evaluation is only conducted through simulation without real dataset. . Sadeghi \emph{et al.} \cite{Sadeghi-ieeewcommletters2019} proposed two AML algorithms  based on a variation of the fast gradient methods (FGMs) \cite{Goodfellow-iclr2015} and tested on the 11-class RadioML \textit{2016.10A} dataset \cite{o2016radio} and with the architecture in \cite{o2016convolutional}. In this paper, we instead consider the much larger RadioML \textit{2018.01A}  dataset \cite{OShea-ieeejstsp2018}, which has 24 classes. 

\section{A Primer on Adversarial Learning}\label{sec:threat_model}


The key target of adversarial machine learning (AML) is to compromise the robustness of classifiers based on neural networks \cite{Moosavi-cvpr2017}. Broadly speaking, there are two types of AML attacks studied in the literature, which are often referred to as \textit{targeted} \cite{Szegedy-2013intriguing} and \emph{untargeted} \cite{Moosavi-cvpr2017}. The former type attempts to find \textit{perturbation vectors} $\mathbf{v}$ that, applied to a given input $\mathbf{x}$, makes the classifier ``steer'' toward a different class than the ground truth $g$. More formally, given a classifier $C$ and a target $t$, the adversary tries to find $\mathbf{x} + \mathbf{v} \sim \mathbf{x}$ such that $C(\mathbf{x} + \mathbf{v}) = t \not= g$. Conversely, untargeted AML attempts to find \textit{universal} perturbation vectors $\mathbf{v}$, through which $C(\mathbf{x} + \mathbf{v}) \not= C(\mathbf{x})$ for most inputs $\mathbf{x}$. To keep the notation consistent with previous work, we will keep the same nomenclature throughout the paper.

\begin{figure}[!h]
    \centering
    \includegraphics[width=0.9\columnwidth]{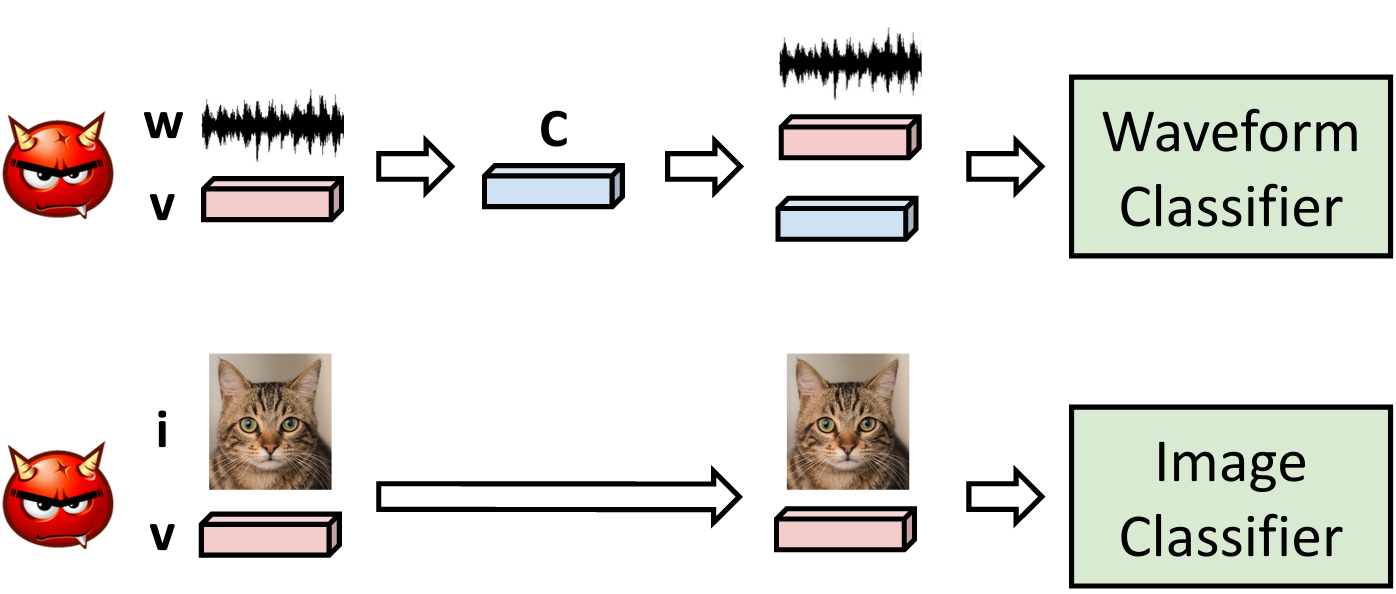}
    \caption{Wireless vs CV-based Adversarial Machine Learning.}\vspace{-0.2cm}
    \label{fig:wireless_vs_cv}
\end{figure}

Figure \ref{fig:wireless_vs_cv} summarizes the differences between AML for Computer Vision (CV) and wireless networking applications. Although very similar in scope and target, there are unique characteristics that make AML in the wireless domain fundamentally different than AML in CV systems. First, CV-based algorithms assume that adversarial and legitimate inputs are received ``as-is'' by the classifier. In other words, if $\mathbf{x}$ is an image and $\mathbf{x} + \mathbf{v}$ is the adversarial input, the classifier will always attempt to classify $\mathbf{x}+\mathbf{v}$ as input. However, due to the wireless channel, we cannot make this assumption in the wireless domain. In short, any adversarial waveform $\mathbf{w} + \mathbf{v}$ will be subject to the additive and multiplicative action of the channel, which can be expressed via a perturbation matrix $\mathbf{C}=(\mathbf{c}_a,\mathbf{c}_m)$ given by the wireless channel, which ultimately makes the classifier attempts to classify the waveform $\mathbf{c}_m (\mathbf{w} + \mathbf{v}) + \mathbf{c}_a$ instead of the $\mathbf{w} + \mathbf{v}$ waveform. The second key difference is that wireless AML has to assume that waveforms cannot be arbitrarily modified, since they have to be decodable at the receiver's side (\textit{i.e.}, if not decodable, the receiver will discard received packets, thus making the attack ineffective). Therefore, the adversary has a critical constraint on the maximum distortion that the joint action $\mathbf{C}$ of the channel and his own perturbation $\mathbf{v}$ can impose to a waveform. Meaning, $\mathbf{c}_m (\mathbf{w} + \mathbf{v}) + \mathbf{c}_a$ still has to be decodable. As we will show in the rest of the paper, an adversary's capability of launching a successful  AML attack will depend on the signal-to-noise ratio (SNR) between the adversary and the receiver. 


\section{MODELING WIRELESS AML}\label{sec:threat_model}


We use boldface upper and lower-case letters to denote matrices and column vectors, respectively. For a vector $\mathbf{x}$, $x_i$  denotes the i-th element, $\norm{\mathbf{x}}_p$ indicates the $l_p$- norm of $\mathbf{x}$, $\mathbf{x}^\top$ its transpose, and $\mathbf{x} \cdot \mathbf{y}$ the inner product of $\mathbf{x}$ and $\mathbf{y}$. For a matrix $\mathbf{H}$, $H_{ij}$ will indicate the (i,j)-th element of $\mathbf{H}$. The notation $\mathbb{R}$ and $\mathbb{C}$ will indicate the set of real and complex numbers, respectively. \smallskip

\begin{figure}[!h]
    \centering
    \includegraphics[width=0.9\columnwidth]{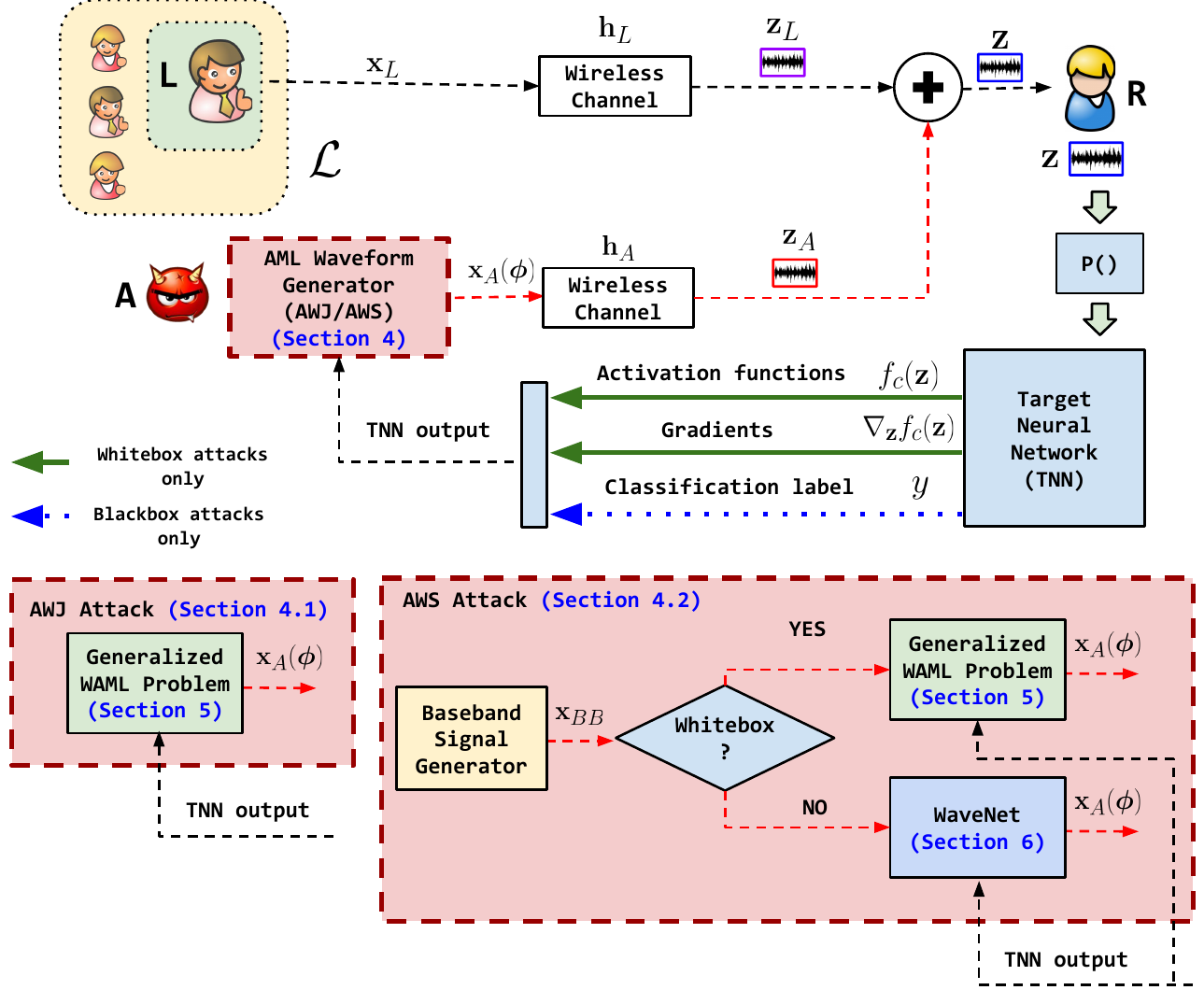}
    \caption{Overview of AML Waveform Jamming (AWJ) and AML Waveform Synthesis (AWS).\vspace{-0.8cm}}
    \label{fig:attacks}
\end{figure}

\textbf{System Model.}~The top portion of Figure \ref{fig:attacks} summarizes our system model, where we consider a receiving node $R$, an attacker node $A$, and a set $\mathcal{L}$ of $N$ legitimate nodes communicating with $R$. We assume that $R$ hosts a target neural network (TNN) used to classify waveforms coming from nodes in $\mathcal{L}$. 

Let $\Lambda>1$ be the number of layers of the TNN, and $\mathcal{C}$ be the set of its classes. We model the TNN as a function $F$ that maps the relation between an input $\mathbf{x}$ and an output $\mathbf{y}$ through a $\Lambda$-layer mapping $F(\mathbf{x}; \mb{\theta}) : \mathbb{R}^{i} \rightarrow \mathbb{R}^{o}$ of an input vector $\mathbf{x} \in \mathbb{R}^{i}$ to an output vector $\mathbf{y} \in \mathbb{R}^{o}$. The mapping happens through $\Lambda$ transformations:
\begin{equation}
    \mathbf{r}_{j} = F_j(\mathbf{r}_{j-1}, \theta_j) \hspace{0.5cm} 0 \le j \le \Lambda,
    \label{eq:dl_layers}
\end{equation}
where $F_j(\mathbf{r}_{j-1}, \theta_j)$ is the mapping carried out by the $j$-th layer. The vector $\boldsymbol{\theta} = \{\theta_1, \ldots, \theta_\Lambda\}$ defines the whole set of parameters of the TNN. We assume the last layer of the TNN is \emph{dense}, meaning that $F_{\Lambda-1}(\mathbf{r}_{j-1}, \theta_j) = \sigma(\mathbf{W}_j \cdot \mathbf{r}_{j-1} + \mathbf{b}_j)$, where $\sigma$ is a softmax activation function, $\mathbf{W}_j$ is the \emph{weight} matrix and $\mathbf{b}_j$ is the \textit{bias} vector. 

We evaluate the activation probabilities of the neurons at the last layer of the TNN. Let $c\in\mathcal{C}$ be a generic class in the classification set of the TNN. We denote $f_c(\mathbf{x})$ as the \textit{activation probability} of the neuron corresponding to class $c$ at the output layer of the TNN when input $\mathbf{x}$ is fed to the TNN. From \eqref{eq:dl_layers}, it follows that
\begin{equation}
    f_c(\mathbf{x}) = F_{\Lambda,c} (\mathbf{r}_{\Lambda-1}, \theta_\Lambda).
\end{equation}

Notice that the mapping $F(\mathbf{x}; \mathbb{\theta}) : \mathbb{R}^{i} \rightarrow \mathbb{R}^{o}$ can be any derivable function, including recurrent networks. By taking as reference \cite{OShea-ieeejstsp2018}, we assume that the input of the TNN is a series of I/Q samples received from the radio interface. We assume that the I/Q samples may be processed through a \textit{processing function} $P()$ before feeding the I/Q samples to the TNN. Common examples of processing functions $P()$ are equalization, demodulation or packet detection. \smallskip




\textbf{Threat Model.}~We assume the adversary $A$ may or may not part of the legitimate set of nodes in $\mathcal{L}$. We call the adversary respectively \emph{rogue} and  \emph{external} in these cases.  We further classify adversarial action based on the \emph{knowledge} that the adversary possesses regarding the TNN. In the first, called in literature \emph{whitebox}, the adversary $A$ has \textit{perfect knowledge} of the TNN activation functions $F_j$, meaning that $A$ has access not only to the output layer $F_\Lambda$ but also to the weight vector $\boldsymbol{\theta}$ (and thus, its gradient as a function of the input). 

In the second scenario, also called \emph{blackbox}, the adversary does not have full knowledge of the TNN, and therefore cannot access gradients. We do assume, however, that the adversary has access to the output of the TNN. Specifically, for any arbitrarily chosen waveform $\mathbf{x}$, the adversary can obtain its label $C(\mathbf{x})=y$ by querying the TNN.
Obtaining the output of the TNN is an issue known as \textit{1-bit feedback learning}, and was studied by Zhang \emph{et al.} in \cite{zhang2016online}.~In our scenario, the adversary could use ACKs or REQs as 1-bit feedback. Specifically, for a given batch $B$ of size $M$, the loss function $L(B)$ can be approximated by observing the number of ACKs or REQs received ($A$) for the current batch and then assign $L(B) = \frac{M - A}{M}$.

The adversary then may choose different strategies to craft adversarial samples over tuples $(\mathbf{x},y)$ obtained from querying the TNN. By referring to prior work, we consider both \textit{targeted} \cite{Szegedy-2013intriguing} and \emph{untargeted} \cite{Moosavi-cvpr2017} attacks. 

\smallskip

\textbf{Wireless Model.~}To be effective, the attacker must be within the transmission range of $R$, meaning that $A$ should be sufficiently close to $R$ to emit waveforms that compromise (to some extent) ongoing transmissions between any node $l\in\mathcal{L}$ and $R$. This scenario is particularly compelling, since not only can $A$ eavesdrop wireless transmissions generated by $R$ (\textit{e.g.}, feedback information such as ACKs or REQs), but also emit waveforms that can be received by $R$ -- and thus, compromise the TNN.

We illustrate the effect of channel action in Figure \ref{fig:attacks}, which can be expressed through well-established models for wireless networks. Specifically, the waveform transmitted by any legitimate node $L\in\mathcal{L}$ and received by $R$ can be modeled as
\begin{equation} \label{eq:legit}
    \bf{z}_{L} = \bf{x}_L \circledast \bf{h}_{L} + \bf{w}_{L},
\end{equation}
\noindent 
where $\bf{x}_L$ represents the waveform transmitted by node $L$, $\circledast$ is the convolution operator; $\bf{h}_{L}$ and $\bf{w}_{L}$ are the fading and noise characterizing the channel between node $L$ and the receiver $R$. 

Similarly, let $\bf{x}_A$ be the waveform transmitted by node $A$, and let $\boldsymbol{\phi}$ be an attack strategy of $A$. 
The attacker utilizes $\boldsymbol{\phi}$ to transform the waveform $\bf{x}_A$ and its I/Q samples. For this reason, the waveform transmitted by $A$ can be written as $\bf{x}_A(\boldsymbol{\phi})$.
For the sake of generality, in this section we do not make any assumption on $\boldsymbol{\phi}$. However, in Section \ref{sec:attacks} we present two examples of practical relevance (\textit{i.e.}, jamming and waveform synthesis) where closed-form expressions for the attack strategy $\boldsymbol{\phi}$ and $\bf{x}_A(\boldsymbol{\phi})$ are derived. The waveform $\bf{z}_{A}$  can be written as
\begin{equation} \label{eq:adversary}
    \bf{z}_{A} = \bf{x}_A(\boldsymbol{\phi}) \circledast \bf{h}_{A} + \bf{w}_{A}.
\end{equation}

Notice that \eqref{eq:legit} and \eqref{eq:adversary} do not assume any particular channel model, nor any particular attack strategy. Therefore, our formulation is very general in nature and able to model a rich set of real-world wireless scenarios. 

In most wireless applications, noise $\mathbf{w}_i$ can be modeled as additive white Gaussian (AWGN). On the contrary, $\mathbf{h}_i$ depends on mobility, multi-path and interference. Although these aspects strongly depend on the application and network scenarios, they are usually assumed to be constant  within the coherence time of the channel, thus allowing us to model $\mathbf{h}_i$ as a Finite Impulse Response (FIR) filter with $K>0$ complex-valued taps.

By leveraging the above properties, the $n$-th component $z_i[n]$ of the  waveform $\mathbf{z}_i$ received from node $i$ can be written as follows:
\begin{equation} \label{eq:za:bb}
    z_i[n] = \sum_{k=0}^{K-1} h_{i_k}[n] x_i[n-k] + w_i[n]
\end{equation}
\noindent
where $x_i[n]$ is the $n$-th I/Q symbol transmitted by node $i$; $h_{i_k}[n]$ and $w_i[n]$ are respectively the $k$-th complex-valued FIR tap and noise coefficients representing the channel effect at time $n$.\smallskip


\section{Wireless AML Attacks} \label{sec:attacks}

With the help of Figure \ref{fig:attacks}, we now introduce the  \textit{AML Waveform Jamming} (Section \ref{sec:jamming}), and \textit{AML Waveform Synthesis} (Section \ref{sec:synt}). 

\vspace{-0.2cm}

\subsection{AML Waveform Jamming (AWJ)} \label{sec:jamming}

In AWJ, an adversary carefully jams the waveform of a legitimate device to confuse the TNN. Since the TNN takes as input I/Q samples, the adversary may craft a jamming waveform that, at the receiver side, causes a slight displacement of I/Q samples transmitted by the legitimate device, thus pushing the TNN towards a misclassification. 


As shown in Figure \ref{fig:attacks}, the waveform $\bf{x}_A$ generated by the attacker node $A$ is aimed at jamming already ongoing transmissions between a legitimate node $L$ and the receiver $R$. In this case, the signal received by $R$ can be written as 
\begin{equation} \label{eq:received_sig}
  \bf{z} = \bf{z}_{A} + \bf{z}_{L}
\end{equation}
\noindent
where $\bf{z}_{A}$ and $\bf{z}_{L}$ are defined in \eqref{eq:legit} and \eqref{eq:adversary}, respectively.\smallskip

\textbf{Attack objectives ad strategies.}~The attacker aims at computing $\bf{x}_A$ so that $C(\mathbf{z}) \neq C(\mathbf{z}_{L})$. Moreover, this attack can be either \textit{targeted} (\textit{i.e.}, $A$ generates jamming waveforms whose superimposition with legitimate signals produce $C(\mathbf{z}) = c_T$, with $c_T$ being a specific target class in $\mathcal{C}$), or \textit{untargeted} (\textit{i.e.}, it is sufficient to obtain $C(\mathbf{z}) \neq c_L$).


In this case, $\mathbf{x}_A(\boldsymbol{\phi})=\boldsymbol{\phi}$. That is, the transmitted waveform corresponds to the actual attack (jamming) strategy. Specifically, we have
\begin{equation} \label{eq:strategy:jamming}
    \mathbf{x}_A(\boldsymbol{\phi}) = (\phi_n^\Re + j \phi_n^\Im)_{n=1,\dots,N_J},
\end{equation}
\noindent
where (i) $a^\Im = \operatorname{Im}(a)$ and $a^\Re = \operatorname{Re}(a)$ for any complex number $a$; and (ii) $N_J>1$ represents the length of the jamming signal in terms of I/Q samples. 
Since $N_J$ might be smaller than the TNN input $N_I$---without losing in generality---we assume that the adversary periodically transmits the sequence of $N_J$ I/Q samples so that they completely overlap with legitimate waveforms and have the same length. However, it is worth to notice that we do not assume perfect superimposition of the jamming signal with the legitimate signal, and thus, adversarial signals are not added in a precise way to the legitimate waveform. \smallskip

\textbf{Undetectability aspects.}~Recall that any invasive attack might reveal the presence of the adversary to the legitimate nodes, which will promptly implement defense strategies \cite{d2016optimal}. For this reason, the adversary aims at generating misclassifications while masquerading the very existence of the attack by computing $\boldsymbol{\phi}$ such that the signal $\bf{z}$ can still be decoded successfully by the receiver (\textit{e.g.}, by keeping the bit-error-rate (BER) lower than a desirable threshold) but yet misclassified. This is because the attacker aims to conceal its presence. If exposed, the receiver might switch to another frequency, or change location, thus making attacks less effective. However, we remark that this constraint can be relaxed if the jammer is not concerned about concealing its presence.
We further assume the attacker has no control over channel conditions (\textit{i.e.}, $\bf{h}_{A}$ and $\bf{w}_{A})$ and legitimate signals (\textit{i.e.}, $\mathbf{z}_L$), meaning that the attacker can control $\mathbf{x}_A(\boldsymbol{\phi})$ only by computing effective strategies $\boldsymbol{\phi}$. \smallskip

\textbf{Addressing non-stationarity.}~An adversary cannot evaluate the channel $\bf{h}_{L}$ in \eqref{eq:legit} -- which is node-specific and time-varying. Also, waveforms transmitted by legitimate nodes vary according to the encoded information, which is usually a non-stationary process. It follows that jamming waveforms that work well for a given legitimate waveform $\bf{z}_{L}$, might not be equally effective for any other $\bf{z}'_{L}\neq\bf{z}_{L}$. Thus, rather than computing the optimal jamming waveform for each $\bf{z}_{L}$, we compute it over a set of consecutive $S$ legitimate input waveforms, also called \textit{slices}.

Let $\rho\in\{0,1\}$ be a binary variable to indicate whether or not the attacker node belongs to the legitimate node set $\mathcal{L}$ (\textit{i.e.}, a rogue node). Specifically, $\rho=1$ if the attacker node is a rogue device and $A\in\mathcal{L}$, $\rho=0$ if the attacker is external (\textit{i.e.}, $A\not\in\mathcal{L}$).
Also, let $c_L$ and $c_A$ be the correct classes of the waveforms transmitted by nodes $L$ and $A$, respectively. \smallskip

\textbf{Untargeted AWJ.}~The adversary aims at jamming legitimate waveforms such that (i) these are misclassified by the TNN; (ii) malicious activities are not detected by the TNN; and (iii) attacks satisfy hardware limitations (\textit{e.g.}, energy should be limited). These objectives and constraints can be formulated through the following \textit{untargeted} AWJ problem \eqref{prob:jamming:untargeted}:
\begin{align}
\underset{\bs{\phi}}{\text{minimize}} & \hspace{0.2cm} \frac{1}{S} \sum_{s=1}^S \left[f_{c_L}({\bf z}_s) + \rho\cdot  f_{c_A}({\bf z}_s) \right] \label{prob:jamming:untargeted} \tag{AWJ-U} \\
    \text{subject to} & \hspace{0.1cm}
    \mathrm{BER}_L(\mathrm{{\bf z}_s)} \leq \mathrm{BER_{max}}, \hspace{0.2cm} s=1,2,\dots,S \label{prob:jamming:untargeted:c1} \tag{C1} \\
    & \hspace{0.1cm}
    \norm{\bf{x}_A(\boldsymbol{\phi})}^2_2 \leq \mathrm{E_{max}}, \hspace{0.2cm} s=1,2,\dots,S \label{prob:jamming:untargeted:c2} \tag{C2}
\end{align}
\noindent
where ${\bf z}_s = \bf{z}_{A} + \bf{z}_{L_s}$, $\bf{z}_{L_s}$ represents the $s$-th slice (or input) of the TNN; Constraint \eqref{prob:jamming:untargeted:c1} ensures that the BER experienced by the legitimate node is lower than the maximum tolerable BER threshold $\mathrm{BER_{max}}$; while \eqref{prob:jamming:untargeted:c2} guarantees that the energy of the jamming waveform does not exceed a maximum threshold $\mathrm{E_{max}}$.
In practise, Constraints  \eqref{prob:jamming:untargeted:c1} and \eqref{prob:jamming:untargeted:c2} ensure that jamming waveforms do not excessively alter the position of legitimate I/Q samples. This is crucial to avoid anti-jamming strategies such as modulation and frequency hopping, among others.
Although Problem \eqref{prob:jamming:untargeted} takes into account Constraints \eqref{prob:jamming:untargeted:c1} and \eqref{prob:jamming:untargeted:c2} only, in Section \ref{sec:optimization} we extend the formulation to larger set of constraints.\smallskip



\textbf{Targeted AWJ.}~By defining $c_T\in\mathcal{C}$ as the target class, we formulate the \textit{targeted} AWJ as 
\begin{align}
\underset{\bs{\phi}}{\text{maximize}} & \hspace{0.2cm} \frac{1}{S} \sum_{s=1}^S \left[ f_{c_T}({\bf z}_s) - \left( f_{c_L}({\bf z}_s) + \rho \cdot f_{c_A}({\bf z}_s) \right) \right] \label{prob:jamming:targeted} \tag{AWJ-T} \\
    \text{subject to} & \hspace{0.1cm}  \mbox{Constraints}~ \eqref{prob:jamming:untargeted:c1},\eqref{prob:jamming:untargeted:c2} \nonumber
\end{align}

When compared to Problem \eqref{prob:jamming:untargeted}, Problem \eqref{prob:jamming:targeted} differs in terms of the objective function. One naive approach would see the adversary maximize the term $\frac{1}{S} \sum_{s=1}^S f_{c_T}({\bf z}_s)$ only. However, the objective of the adversary is to produce misclassifications, so the adversary should try to reduce the activation probability of the jammed class $c_L$ and adversarial class $c_A$, while maximizing the activation probability for the target class $c_T$. It is expected that the TNN has high accuracy and by simply maximizing $\frac{1}{S} \sum_{s=1}^S f_{c_T}({\bf z}_s)$ does not necessarily mean that the TNN would not be able to still correctly classify transmissions from the legitimate device $L$ (i.e., the activation probability $f_{c_L}$ might still be high). 

Let us provide a simple yet effective example. Assume that the attacker is external ($\rho=0$), $\frac{1}{S} \sum_{s=1}^S f_{c_T}({\bf z}_{L_s}) = 0.1$ and $\frac{1}{S} f_{c_L}({\bf z}_{L_s}) = 0.9$.
Let us consider the case where the adversary computes $\bs{\phi}$ such that the term $\frac{1}{S} \sum_{s=1}^S f_{c_T}({\bf z}_s)$ only is maximized. A reasonable outcome of this optimization problem is that $\bs{\phi}$ is such that $\frac{1}{S} \sum_{s=1}^S f_{c_T}({\bf z}_s) = 0.4$ and $\frac{1}{S} \sum_{s=1}^S f_{c_L}({\bf z}_s) = 0.6$. In this case, it is easy to notice that input waveforms are still classified as belonging to class $c_L$.
A similar argument can be made for term $\rho f_A({\bf z}_s)$ when $\rho = 1$ (\textit{i.e.}, the attacker is a rogue node).

In other words, to effectively fool the TNN, the attacker must generate waveforms that (i) suppress features of class $c_L$; (ii) mimic those of class $c_T$; and (iii) hide features of the attacker's class $c_A$. These objectives can be formulated via the objective function in Problem \eqref{prob:jamming:targeted}.\vspace{-0.2cm}

\subsection{AML Waveform Synthesis (AWS)} \label{sec:synt}

In this attack -- illustrated in the bottom-right side of Figure \ref{fig:attacks} -- an adversary  $A$ transmits synthetic waveforms trying to imitate features belonging to a target class $c_T$. In contrast to the AWJ, in this case $\bf{z} = \bf{z}_{A}$ and synthetic waveforms $\bf{x}_A(\bs{\phi})$ are generated so that $C(\mathbf{z}) = c_T$ and the waveform received by node $R$ is still intelligible. By definition, this attack is \textbf{targeted} only.

Let $c_T\in\mathcal{C}$ be the target class. The (targeted) AWS problem (AWS) is formulated as
\begin{align}
\underset{\bs{\phi}}{\text{maximize}} & \hspace{0.2cm} \frac{1}{S} \sum_{s=1}^S \left[ f_{c_T}({\bf z}_{A_s}) - \rho f_{c_A}({\bf z}_{A_s}) \right] \label{prob:aws} \tag{AWS} \\
    \text{subject to} & \hspace{0.1cm}  \mbox{Constraints}~ \eqref{prob:jamming:untargeted:c1}, \eqref{prob:jamming:untargeted:c2} \nonumber
\end{align}


This attack maps well to scenarios such as radio fingerprinting, where a malicious device aims at generating a waveform embedding impairments that are unique to the target legitimate device \cite{restuccia2019deepradioid}. In other words, the attacker \textit{cannot generate random waveforms as in the AWJ}, but should transmit waveforms that contain decodable information. To this end, FIR filters are uniquely positioned to address this issue. More formally, a FIR is described by a finite sequence $\bs \phi$ of $M$ \textit{filter taps}, \textit{i.e.}, $\bs \phi=(\phi_1,\phi_2,\dots,\phi_M)$. 
For any input $\mf x \in \mc X$, the filtered $n$-th element $\hat{x}[n] \in \hat{\mf x}$ can be written as  
\begin{equation} \label{eq:FIR:general}
    \hat{x}[n] = \sum_{m=0}^{M-1} \phi_m x[n-m]
\end{equation}

It is easy to observe that by using FIRs, \textit{the adversary can manipulate the position in the complex plane of the transmitted I/Q symbols}. By using complex-valued filter taps, \textit{i.e.}, $\phi_m\in \mb C$ for all $m=0,1,\dots,M-1$, Eq. \eqref{eq:FIR:general} becomes:

\begin{align} 
\hat{x}[n] & = \sum_{m=0}^{M-1} (\phi^\Re_m +j\phi^\Im_m) (x^\Re[n-m]+jx^\Im[n-m]) \nonumber \\
& = \hat{x}^\Re[n] +  j \hat{x}^\Im[n]\label{eq:FIR:complex}
\end{align}

For example, to rotate all I/Q samples by $\theta=\pi/2$ radiants and halve their amplitude, we may set $\phi_1 = \frac{1}{2} \exp^{j\frac{\pi}{2}}$ and $\phi_k = 0$ for all $k>1$. Similarly, other complex manipulations can be obtained by fine-tuning filter taps. It is clear that complex FIRs can be effectively used by the attacker node to fool the TNN through AWS attacks. 

By using a FIR $\bs \phi$ with $M$ complex-valued taps, the waveform $\mathbf{x}_A(\boldsymbol{\phi})$ transmitted by the attacker can be written as
\begin{equation}  \label{eq:strategy:synth}
    \mathbf{x}_A(\boldsymbol{\phi}) = \mathbf{x}_{BB} \circledast \boldsymbol{\phi}
\end{equation}
\noindent
where $\mathbf{x}_A(\boldsymbol{\phi})=(x_A[n](\boldsymbol{\phi}))_{n=1,\dots,N_I}$, $x_A[n](\boldsymbol{\phi})$ is computed as in \eqref{eq:FIR:complex}, $\mathbf{x}_{BB} = (x_{BB}[n])_{n=1,\dots,N_I}$ is an intelligible signal (e.g., a portion of a WiFi packet) and $\boldsymbol{\phi}=(\phi_n^\Re + j \phi_n^\Im)_{n=1,\dots,N_I}$ is the FIR used to generate a synthetic waveform.

\section{Generalized WAML Problem (GWAP)}\label{sec:optimization}

Notice that Problems \eqref{prob:jamming:untargeted}, \eqref{prob:jamming:targeted} and \eqref{prob:aws} are similar in target. Thus,  we propose the following Generalized Wireless AML problem (GWAP) formulation
\begin{align}
\underset{\boldsymbol{\phi}}{\text{maximize}} & \hspace{0.2cm} \sum_{s=1}^S \sum_{c\in\mathcal{C}} \omega_c f_c({\bf z}_s)  \label{prob:gap} \tag{GWAP} \\
    \text{subject to} & \hspace{0.1cm}  \mathbf{g}(\mathbf{z}_s) \leq 0, \hspace{0.1cm} s=1,...,S
\end{align}
\noindent
where $\mathbf{g}(\mathbf{z})=(g_1(\mathbf{z}),\dots,g_G(\mathbf{z}))^\top$ is a generic set of constraints that reflect BER, energy and any other constraint that the attack strategy $\boldsymbol{\phi}$ must satisfy (\textit{e.g.}, upper and lower bounds); and $\omega_c$ takes values in $\{-\rho,-1,0,1,\rho\}$ depending on the considered attack. As an example, Problem \eqref{prob:jamming:targeted} has $\omega_{c_T} = 1$, $\omega_{c_L} = -1$, $\omega_{c_A} = -\rho$ and $\omega_c = 0$ for all $c\neq c_L,c_T,c_A$.

Problem \eqref{prob:gap} is non trivial since (i) the functions $f_c$ have no closed-form and depend on millions of parameters; (ii) both the objective and the constraints are highly non-linear and non-convex; (iii) it is not possible to determine the convexity of the problem. Despite the above challenges, in whitebox attacks \textit{the adversary has access to the gradients of the TNN} (Figure \ref{fig:attacks}).   
In the following, we show how an attacker can effectively use gradients 
to efficiently compute AML attack strategies. 
It is worth mentioning that our whitebox algorithms, similar to the fast gradient sign method (FGSM) \cite{huang2017adversarial}, use gradients to generate adversarial outputs. Despite being similar, FGSM can compute adversarial examples tailored for a specific input and a specific channel condition only. Conversely, as explained in Section \ref{sec:jamming}, under "Addressing non-stationarity", our algorithms take into account \textit{multiple inputs} to find a single FIR filter that can synthesize adversarial inputs for multiple channel conditions, thus resulting more general and practical than FGSM-based approaches.

From \eqref{eq:received_sig}, the input of the TNN is $\mathbf{z}=\mathbf{z}_A + \mathbf{z}_L$. Since $\mathbf{z}_L$ cannot be controlled by the attacker node, we have $f_c(\mathbf{z})=f_c(\mathbf{z}_A)$. Figure \ref{fig:attacks} shows that the TNN provides the gradients $\nabla_{\mathbf{z}} f_c(\mathbf{z})$, hence the attacker can compute the gradients $\nabla_{\boldsymbol{\phi}} f_c(\mathbf{z})$ of the activation probability corresponding to the $c$-th class of the TNN with respect to the attacker's strategy $\boldsymbol{\phi}$ by using the well-known chain rule of derivatives. Specifically, the gradients are
\begin{equation} \label{eq:chain:vector}
    \nabla_{\boldsymbol{\phi}}f_c(\mathbf{z}) = J_{\boldsymbol{\phi}}(\mathbf{z})^\top \cdot \nabla_{\mathbf{z}}f_c(\mathbf{z})
\end{equation}
\noindent
where $J_{\boldsymbol{\phi}}(\mathbf{z})$ is the $N_I \times M$ Jacobian matrix of the input $\mathbf{z}$ with respect to the attacker's strategy $\boldsymbol{\phi}$, $\top$ is the transposition operator, and $\cdot$ stands for matrix dot product. 

We define the input of the TNN  as a set of $N_I$ consecutive I/Q samples, \textit{i.e.}, $\mathbf{z} = (z[n])_{n=0,\dots,N_I-1}$, where $z_n\in\mathbb{C}$ for all $n=0,\dots,N_I-1$. The attacker's waveform is defined as a sequence of $M$ complex numbers, \textit{i.e.}, $\mathbf{x}_A(\boldsymbol{\phi})=(x_{A}[m](\boldsymbol{\phi}))_{m=0,\dots,M-1}$ whose values depend on the attack strategy $\boldsymbol{\phi}$. With this information at hand, we observe the gradient $\nabla_{\boldsymbol{\phi}}f_c(\mathbf{z})$ has dimension $2M\times 1$, while the gradients with respect to real and imaginary parts of the $m$-component are respectively
\begin{align} 
    \frac{\partial f_c(\bf z) }{\partial \phi_m^\Re} & \!\!=\!\! \sum_{n = 1}^{N_I} \left ( \frac{\partial f_c(\bf z)}{\partial z^\Re[n]} \frac{\partial z^\Re[n] }{\partial \phi_m^\Re} + \frac{\partial f_c(\bf z)}{\partial z^\Im[n]} \frac{\partial z^\Im[n] }{\partial \phi_m^\Re} \right) \label{eq:par:re}\\
    \frac{\partial f_c(\bf z) }{\partial \phi_m^\Im} & \!\!=\!\! \sum_{n = 1}^{N_I} \left ( \frac{\partial f_c(\bf z)}{\partial z^\Re[n]} \frac{\partial z^\Re[n] }{\partial \phi_m^\Im} + \frac{\partial f_c(\bf z)}{\partial z^\Im[n]} \frac{\partial z^\Im[n] }{\partial \phi_m^\Im} \right). \label{eq:par:im}
\end{align}

\subsection{Gradients Computation}

We remark that while the AWJ generates waveforms that mimic noise on the channel and target already ongoing transmissions between legitimate nodes, the AWS aims at creating synthetic waveforms when no other node is occupying the wireless channel. Therefore, the two attacks require different attacks strategies $\boldsymbol{\phi}$ which will inevitably result in different values of \eqref{eq:par:re} and \eqref{eq:par:im}. Thus, we discuss the implementation details of AWJ and AWS attacks and derive the corresponding closed-form expressions for the partial derivatives in \eqref{eq:par:re} and \eqref{eq:par:im}. \smallskip

\textbf{AML Waveform Jamming}.~Here, the adversary is not required to transmit intelligible or standard-compliant waveforms. Therefore, $\mathbf{x}_A(\boldsymbol{\phi})$ is defined in \eqref{eq:strategy:synth}. Since $\boldsymbol{\phi}$ is the only variable the attacker can control, $\frac{\partial z^{Z'}[n] }{\partial \phi^{Z''}_m} = \frac{\partial z_{A}^{Z'}[n] }{\partial \phi^{Z''}_m}$, where $Z'$ and $Z''$ can be either $\Re$ or $\Im$ to identify real and imaginary part, respectively. Accordingly, from \eqref{eq:za:bb} we have
\begin{equation} \label{eq:partial:jam}
    \frac{\partial z^{Z'}[n] }{\partial \phi^{Z''}_m} = h_{A_{n-m}}[n]
\end{equation}

By substituting \eqref{eq:partial:jam} into \eqref{eq:par:re} and \eqref{eq:par:im}, the attacker can calculate gradients that will be used to compute an efficient jamming solution in Section \ref{sec:gradient_based}.


\textbf{AML Waveform Synthesis}.~In this attack, the optimization variables are the FIR taps and the attacker's waveform $\mathbf{x}_A(\boldsymbol{\phi})$ is defined in \eqref{eq:strategy:synth}. For this reason, gradients can be computed as follows:
\begin{equation} \label{eq:partial:syn}
    \frac{\partial z^{Z'}[n] }{\partial \phi^{Z''}_m} = \sum_{k=0}^{K-1} h_{A_k}[n] \left( \sum_{m=0}^{M-1} x_{BB}[n-m-k] \right)
\end{equation}

\subsection{Gradient-based Optimization Algorithm} \label{sec:gradient_based}

Now we present a general solution to Problem \ref{prob:gap} which leverages the availability of gradients \eqref{eq:par:re}, \eqref{eq:par:im}, \eqref{eq:partial:jam} and \eqref{eq:partial:syn} to compute an effective attack strategy $\boldsymbol{\phi}$. 

First, we relax the
constraints $g_i(\cdot)$ through Lagrangian Relaxation \cite{bertsekas2014constrained}. Specifically, we define the \textit{augmented Lagrangian}
\begin{align} \label{eq:lagrangian}
    L(\boldsymbol{\phi},\boldsymbol{\lambda}) & = \sum_{s=1}^S \left( \sum_{c\in\mathcal{C}} \omega_c f_c({\bf z}_s) -  \boldsymbol{\lambda}_s^\top \mathbf{g}(\mathbf{z}_s) - \frac{\rho}{2} ||\mathbf{g}(\mathbf{z}_s)||^2_2 \right)
\end{align}
\noindent
where $\boldsymbol{\lambda}_s=(\lambda_{0,s},\dots,\lambda_{G,s})^\top$, $\lambda_{G,s}\geq 0$, $\boldsymbol{\lambda}=(\boldsymbol{\lambda}_1,\dots,\boldsymbol{\lambda}_S)$, and $\rho>0$ is a fixed-step size to regulate the convergence speed of the algorithm \cite{bertsekas2014constrained}. By using Lagrangian duality, an approximated solution to Problem \eqref{prob:gap} can be found by the following iterative process
\begin{align}
    \boldsymbol{\phi}^{(t)} & = \argmax_{\boldsymbol{\phi}} L(\boldsymbol{\phi},\boldsymbol{\lambda}^{(t-1)}) \label{eq:iter:phi}\\ 
    \boldsymbol{\lambda}_s^{(t)} & = \max\{0,\boldsymbol{\lambda}_s^{(t-1)} + \gamma_t \mathbf{g}(\mathbf{z}_s)\}
\end{align}
\noindent
where $t$ represents the iteration counter and $\gamma_t$ is a decreasing step-size such that $\sum_t \gamma_t = \infty$ and $\sum_t \gamma^2_t < \infty$ \cite{bertsekas2014constrained}.

We solve \eqref{eq:iter:phi} via the Non-linear Conjugate Gradient (NCG) method \cite{hestenes1952methods}. To compute a solution at each iteration $t$, we define the gradient of $L(\boldsymbol{\phi},\boldsymbol{\lambda}^{(t-1)})$ as a function of the attack strategy $\mathbf{\phi}$:
\begin{align}
    \nabla_{\mathbf{\phi}}L(\boldsymbol{\phi},\boldsymbol{\lambda}^{(t-1)}) & = \sum_{s=1}^S \sum_{c\in\mathcal{C}} \omega_c \nabla_{\boldsymbol{\phi}}f_c(\mathbf{z}_s) \nonumber \\
    & - \boldsymbol{\lambda}_s^{(t-1)\top} \nabla_{\boldsymbol{\phi}}\mathbf{g}(\mathbf{z}_s) - \rho J_\mathbf{g}^\top(\boldsymbol{\phi})\cdot\mathbf{g}(\mathbf{z}_s)
\end{align}
\noindent
with $\nabla_{\boldsymbol{\phi}}f_c(\mathbf{z}_s)$ being computed in \eqref{eq:chain:vector}, $\nabla_{\boldsymbol{\phi}}\mathbf{g}(\mathbf{z}_s)$ and $J_\mathbf{g}^\top(\boldsymbol{\phi})$ being the gradient and Jacobian matrix of the functions $\mathbf{g}$ with respect to $\boldsymbol{\phi}$, respectively. We omit the NCG-based solution, and refer the interested reader to \cite{bertsekas2014constrained,hestenes1952methods} for a theoretical background of the algorithm.

\section{Blackbox Optimization: FIRNet}\label{sec:learning_approach}


The core objective of \textit{FIRNet} is to hack the TNN \emph{without} requiring to have a copy of the TNN. To this end, we leverage the feedback from the TNN to carefully transform the input via a series of \textit{finite impulse response (FIR) convolutional layers}, which to the best of our knowledge are conceived for the first time in this paper. 

\begin{figure}[!h]
    \centering
    \includegraphics[width=0.9\columnwidth]{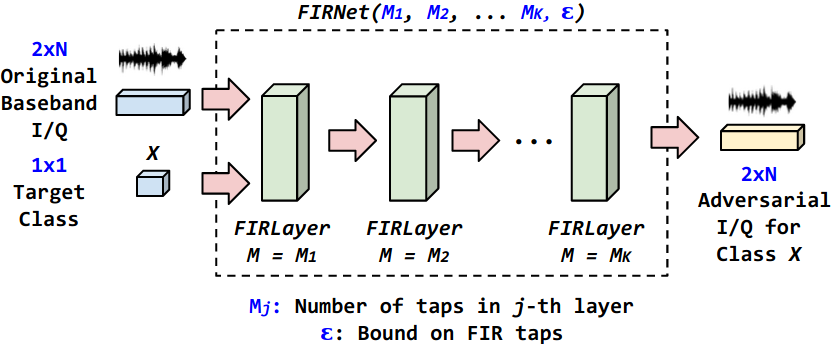}
    \caption{The \textit{FIRNet} Architecture.\vspace{-0.7cm}}
    \label{fig:FIRNet}
\end{figure}

Figure \ref{fig:FIRNet} shows at a high level the architecture of \textit{FIRNet}. In a nutshell, the  ultimate target of \textit{FIRNet} is to take as input a number of I/Q samples generated by the adversary's wireless application and a \emph{target class} that the  and ``perturbate'' them through a series of consecutive \textit{FIRLayers}. The key intuition is that FIR operations are easily implementable in software and in hardware, making the complexity of \textit{FIRNet} scalable. Moreover, an FIR can be implemented using one-dimensional (1D) layers in Keras. Thus, \textit{FIRNet} is fully GPU-trainable and applicable to many different applications  beside the ones described in this paper. More formally, by defining \(x^R\), \(x^I\) the real and imaginary components of an I/Q signal, and \(\phi^R\), \(\phi^I\) the real and imaginary components of the FIR, a \emph{FIRLayer} manipulates an input as follows:

\begin{align}
	y[n] = \sum^{N-1}_{i=0} (\phi^R_i + j\phi^I_i)(x^R[n-i]+jx^I[n-i]),
\end{align}

Before training, the \textit{FIRLayer}'s weights are initialized such that \(\phi_0 = 1\) and \(\{\phi_i\} = 0, i > 0\). This initialization in essence represents an identity vector, which returns unchanged input values. The reason why we consider this particular initialization rule is to preserve the shape and content of input waveforms in the first few training epochs. This way FIRNet updates weights iteratively without irremediably distorting input waveforms. 

\subsection{FIRNet Training Process}

We first provide a brief background on traditional adversarial networks, and then we formalize the \textit{FIRNet} training process. 

Generative adversarial networks (GANs) are composed by a generator $G$ and a discriminator $D$. Both $G$ and $D$ are trained to respectively learn (i) the data distribution and (ii) to distinguish samples that come from the training data rather than $G$.~To this end, the generator builds a mapping function parametrized with $\theta_g$ from a prior noise distribution $p_z$ as $G(z; \theta_g)$, while the discriminator $D(x; \theta_d)$, parametrized with $\theta_d$ parameters,  outputs a single scalar representing the probability that $x$ came from the training data distribution $p_x$ rather than the generator $G$. Therefore, $G$ and $D$ are both trained simultaneously in a minmax problem, where the target is to find the $G$ that minimizes $\log{1-D(G(z))}$ and the $D$ that minimizes $\log{D(\mathbf{x})}$. More formally: 
\begin{equation}
\underset{\bs G}{\text{min}}\ \underset{\bs D}{\text{max}}  \hspace{0.2cm}  \mathbb{E}_{\mathbf{x}\sim p_x} \log(D(\mathbf{x})) + \mathbb{E}_{\mathbf{z}\sim p_z} \log(1-D(G(\mathbf{z}))))
\label{eq:gans}
\end{equation}

Although \emph{FIRNet} is at its core an adversarial network, there are a number of key aspects that set \emph{FIRNet} apart from existing GANs. First, in our scenario $D$ has already been trained and thus is not subject to any modification during the $G$ training process. Second, GANs assume that $D$ is a binary discriminator (\textit{i.e.}, "fake" vs ``authentic'' response). This  is not the case in our problem, since $D$ has a softmax output (\textit{i.e.}, multiclass). Third, GANs take as input a noise vector, whereas here we need to take baseband I/Q samples as inputs. Fourth, as shown in Equation \ref{eq:gans}, the minmax problem solved by GANs is unconstrained, while the GWAP problem in Section \ref{sec:optimization} is instead constrained. Fifth, GANs assume stationarity, which is not entirely the case in the wireless domain. Finally, to really implement a ``blackbox'' attack, we cannot assume that the waveform produced by \emph{FIRNet} will be used by the target network without further processing (\textit{e.g.}, demodulation), which is instead assumed in traditional GANs.

\begin{figure}[!h]
    \centering
    \includegraphics[width=0.9\columnwidth]{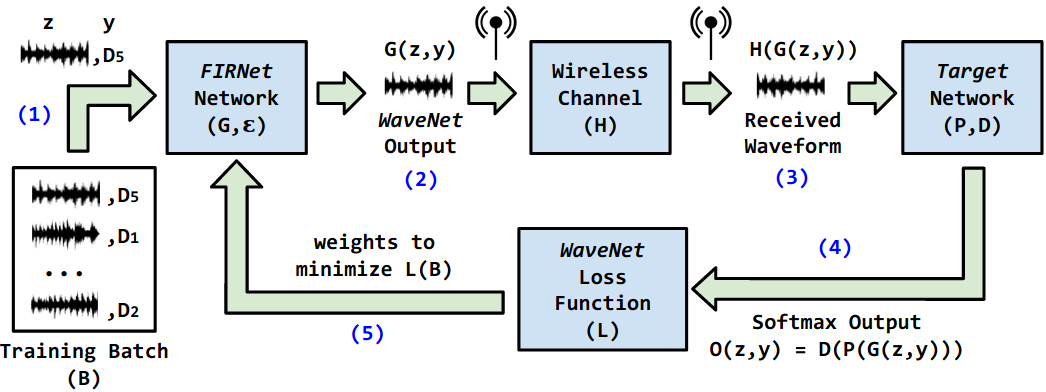}
    \caption{The \textit{FIRNet} Training Loop.\vspace{-0.2cm}}
    \label{fig:FIRNet_training}
\end{figure}

For the above reasons, we devise a brand-new training strategy shown in Figure \ref{fig:FIRNet_training}. In a nutshell, we aim to train a generator function $G$ able to imitate \emph{any} device the target network $D$ has been trained to discriminate and with \emph{any} baseband waveform of interest. As in previous work \cite{restuccia2019deepradioid}, to limit the FIR action to a given scope we model the constraint \eqref{prob:jamming:untargeted:c1} in Problem \eqref{prob:jamming:untargeted} as a box constraint where each I/Q component of the FIR is constrained within $[-\epsilon,\epsilon]^2$, for any small $\epsilon>0$.

First, the adversary generates a waveform training batch $B$ (step 1), where waveforms are generated according to the wireless protocol being used. For example, if WiFi is the wireless protocol of choice, each waveform could be the baseband I/Q samples of a WiFi packet that the adversary wants to transmit. To each waveform $z$ in the batch, the adversary assigns an \textit{embedded label} $y$, which is selected randomly among the set of devices that the adversary wants to imitate. Notice that the adversary does not need to know exactly the number of devices in the network. This set is then fed to \textit{FIRNet} which generates a training output $G(z, y, \epsilon)$ (step 2), where $\epsilon$ is the constraint of the weight of the FIRLayers as explained earlier. 

The waveform produced by \emph{FIRNet} is then transmitted over the air and then received as a waveform $H(G(z, y, \epsilon))$ (step 3). It is realistic to assume that the device could pre-process the waveform before feeding it to the target network, \textit{e.g.}, to extract features in the frequency domain \cite{restuccia2019deepradioid,12_vo2016fingerprinting}. Thus, the softmax output of the target network is modeled as $O(z, y) = D(P(H(G(z, y, \epsilon))))$. We assume that the adversary does not have access in any way to $D$ and $P$, but only to the softmax output. The adversary can thus minimize the following loss:

\begin{equation}
L(B) = -\sum_{(z,y) \in B}\sum_{t=1}^M \mathbb{I}\{ t = y \} \cdot \log(O_t(z, y))
\label{eq:gans}
\end{equation}

\noindent where $M$ is the number of devices, $\mathbb{I}\{\cdot\}$ is a binary indicator function, and $O_t$ is the softmax output for target class $t$. The adversary can then minimize $L(B)$ using stochastic gradient descent (SGD) or similar algorithms.




\section{Experimental Results}\label{sec:exp_res}

We first describe the datasets and learning architectures in Section \ref{sec:datasets}, followed by the results for AWJ (Section \ref{sec:awj_results}),  AWS (Section \ref{sec:aws_results}) and FIRNet (Section \ref{sec:over_the_air}).\vspace{-0.2cm}

\subsection{Datasets and Learning Architectures}\label{sec:datasets}

\subsubsection{Radio Fingerprinting}

We consider (i) a dataset of 500 devices emitting IEEE 802.11a/g (WiFi) transmissions;  and (ii) a dataset of 500 airplanes emitting Automatic Dependent Surveillance -- Broadcast (ADS-B) beacons\footnote{Due to stringent contract obligations, we cannot release these datasets to the community. We hope this will change in the future.}. ADS-B is a surveillance transmission where an aircraft determines its position via satellite navigation. For the WiFi dataset, we demodulated the transmissions and trained our models on the derived I/Q samples. To demonstrate the generality of our AML algorithms, the ADS-B model was instead trained on the unprocessed I/Q samples. We use the CNN architecture in \cite{5_8466371}, where the input is an I/Q sequence of length $288$, followed by two convolutional layers (with ReLu and 2x2 MaxPool) and two dense layers of size 256 and 80. We refer to the above CNN models as \textbf{RF-W (WiFi) and RF-A (ADS-B) TNN architectures}.

\subsubsection{Modulation Classification (MC)}

We use the RadioML 2018.01A dataset, \textbf{publicly available for download at \url{http://deepsig.io/datasets}}. The dataset is to the best of our knowledge the largest modulation dataset available, and includes 24 different analog and digital modulations generated with different levels of signal-to-noise ratio (SNR). Details can be found in \cite{OShea-ieeejstsp2018}. For the sake of consistency, we also consider the neural network introduced in Table III of \cite{OShea-ieeejstsp2018}, which presents 7 convolutional layers each followed by a MaxPool-2 layer, finally followed by 2 dense layers and 1 softmax layer. The dataset contains 2M examples, each 1024 I/Q samples long. In the following, this model will be referred to as the \textbf{MC TNN architecture}. We considered the same classes shown in Figure 13 of \cite{OShea-ieeejstsp2018}. Confusing classes in Fig. \ref{fig:jam_any_mod_lowsnr_matrix} ($\epsilon=0.2$) of our paper and Figure \cite{OShea-ieeejstsp2018} in  are the same (\textit{i.e.}, mostly M-QAM modulations). Notice that $\epsilon=0$ corresponds to zero transmission power (\textit{i.e.}, no attack).\smallskip

\subsubsection{Data and Model Setup}~For each architecture and experiment, we have extracted two distinct datasets for \textit{testing} and \textit{optimization} purposes. The optimization set is used to compute the attack strategies $\boldsymbol{\phi}$ as shown in Sections \ref{sec:attacks} and \ref{sec:optimization}. The computed $\boldsymbol{\phi}$ are then applied to the testing set and then fed to the TNN.  To understand the impact of channel conditions, we simulate a Rayleigh fading channel with AWGN noise $\mathbf{h}_A$ that affects all waveforms that node $A$ transmits to node $R$. We consider high and low SNR scenarios with path loss equal to $0$dB and $20$dB, respectively. Moreover, we also consider a baseline case with no fading.

\subsubsection{Model Training}\label{sec:training_models}

To train our neural networks, we use an $\ell_2$ regularization parameter $\lambda=0.0001$. We also use an Adam optimizer with a learning rate of $l = 10^{-4}$ and categorical cross-entropy as a loss function. All architectures are implemented in Keras. \textbf{The source code used to train the models is free and available to the community for download at \url{https://github.com/neu-spiral/RFMLS-NEU}.}

\subsection{AML Waveform Jamming (AWJ) Results}\label{sec:awj_results}

In AWJ, the adversary aims at disrupting the accuracy of the TNN by transmitting waveforms of length $N_J$ and of maximum amplitude $\epsilon>0$, to satisfy Constraint \eqref{prob:jamming:untargeted:c2} and keep the energy of the waveform limited. 

\subsubsection{Untargeted AWJ (U-AWJ)} 

\begin{figure}[!h]
    \centering
    \includegraphics[width=0.9\columnwidth]{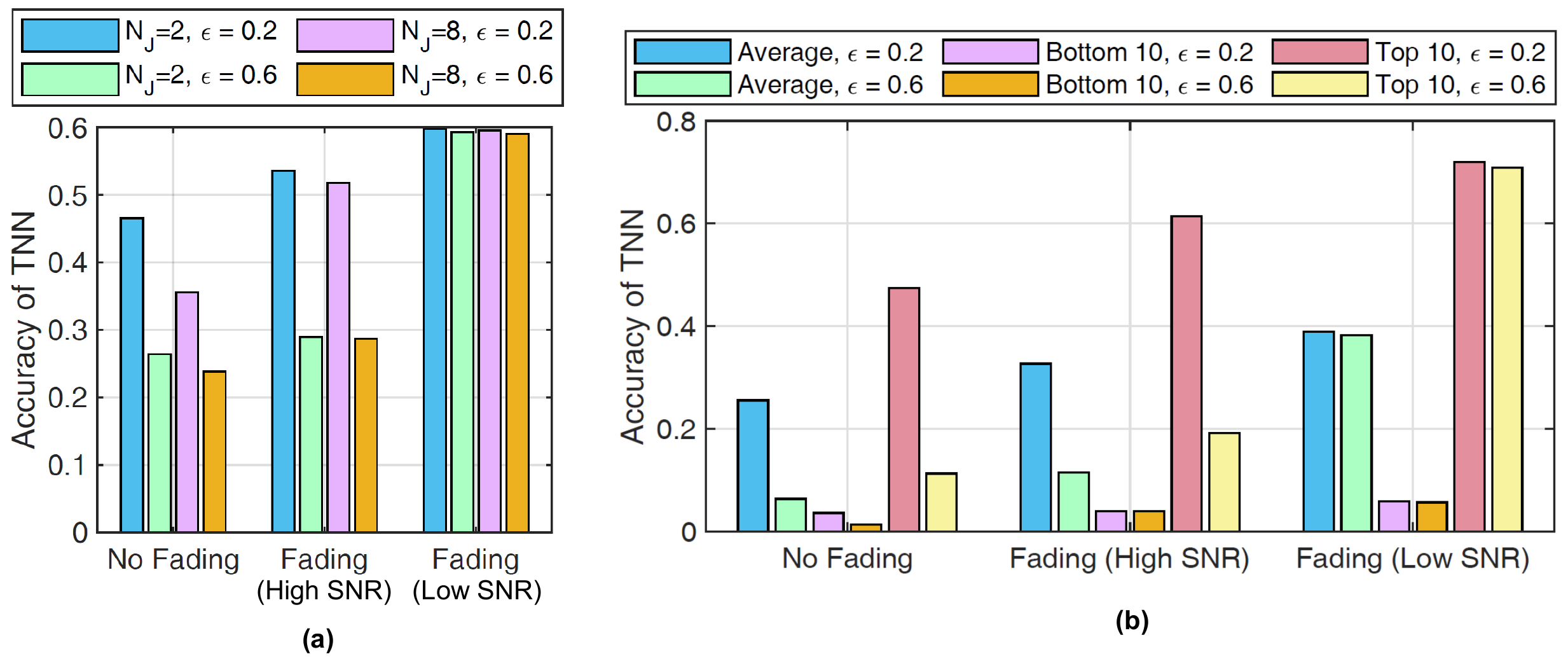}
    \caption{\label{fig:jam_any_mod_bars} Accuracy of (a) MC TNN (originally $60\%$) and (b) RF-W TNN (originally $40\%$) under the AWJ-U attack for different jamming lengths and $\epsilon$ values.\vspace{-0.2cm}} 
\end{figure}

Figure \ref{fig:jam_any_mod_bars}(a) shows the accuracy of the MC TNN (original accuracy of $60\%$) under the AWJ-U attack, for different channel conditions $\mathbf{h}_A$, jamming waveform length $N_J$ and $\epsilon$ values. Figure \ref{fig:jam_any_mod_bars} shows that the adversary always reduces the accuracy of the TNN even when $N_J$ and $\epsilon$ are small. We notice that high SNR fading conditions allow the adversary to halve the accuracy of the TNN, while the best performance is achieved in no-fading conditions where the attacker can reduce the accuracy of the TNN by a 3x factor. 

\begin{figure}[h!]
\centering
  \begin{minipage}[b]{0.60\columnwidth}
    \includegraphics[width=\columnwidth]{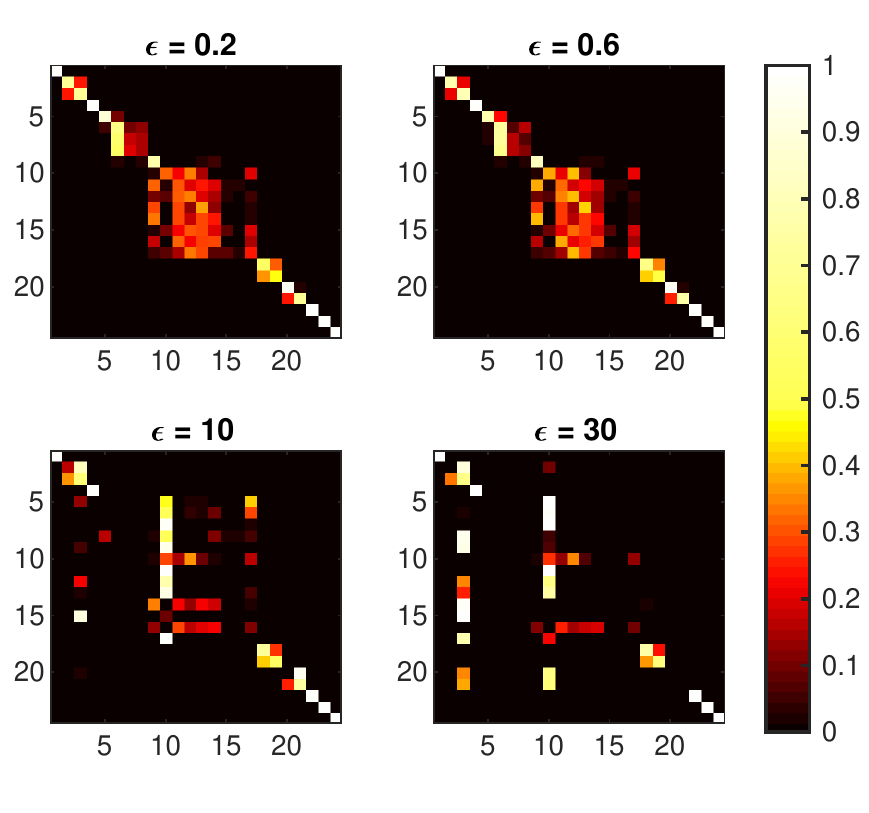}
    \caption{\label{fig:jam_any_mod_lowsnr_matrix} Confusion matrix of MC TNN under the AWJ-U attack in low SNR regime for different $\epsilon$ values.}
  \end{minipage}
  \hspace{0.1cm}
   \begin{minipage}[b]{0.33\columnwidth}
    \centering
    \includegraphics[width=\columnwidth]{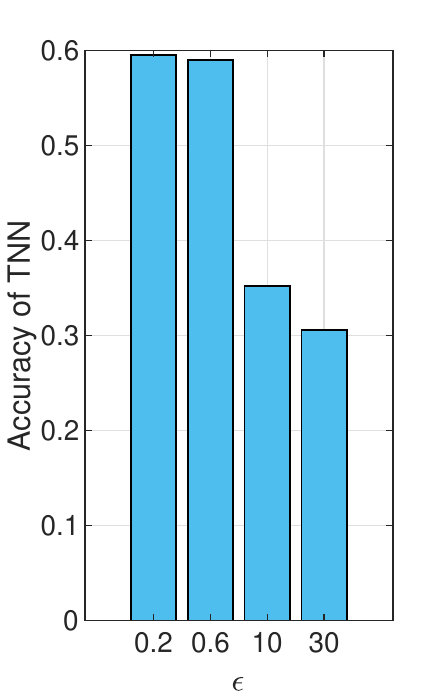}
    \vspace{-0.5cm}
    \caption{\label{fig:jam_any_mod_lowsnr_bars} Accuracy of MC TNN in Fig. \ref{fig:jam_any_mod_lowsnr_matrix} (originally $60\%$).}
  \end{minipage}
\end{figure}

Figures \ref{fig:jam_any_mod_lowsnr_matrix} and \ref{fig:jam_any_mod_lowsnr_bars} show the confusion matrices and the corresponding accuracy levels of the AWJ-U attack to the MC TNN model in the low SNR regime. Here, increasing $\epsilon$ also increases the effectiveness of the attack, demonstrated by the presence of high values outside the main diagonal of the confusion matrix.

Figure \ref{fig:jam_any_mod_bars}(b) shows the accuracy of the RF-W TNN for different attack strategies, constraints and fading conditions. To better understand the impact of AWJ-U, we have extracted the 10 least (\textit{i.e.}, \textit{Bottom 10}) and most (\textit{i.e.}, \textit{Top 10}) classified devices out of the 500 devices included in the WiFi dataset. 
Interestingly, AWJ-U attacks are extremely effective when targeting the top devices. In some cases, the attacker can drop the accuracy of these devices from $70\%$ to a mere $20\%$ in the high SNR regime. Since the bottom 10 devices are classified with a low accuracy already, it takes minimal effort to alter legitimate waveforms and activate other classes. 


\subsubsection{Targeted AWJ (AWJ-T)} 

\begin{figure}[!h]
    \centering
    \includegraphics[width=0.9\columnwidth]{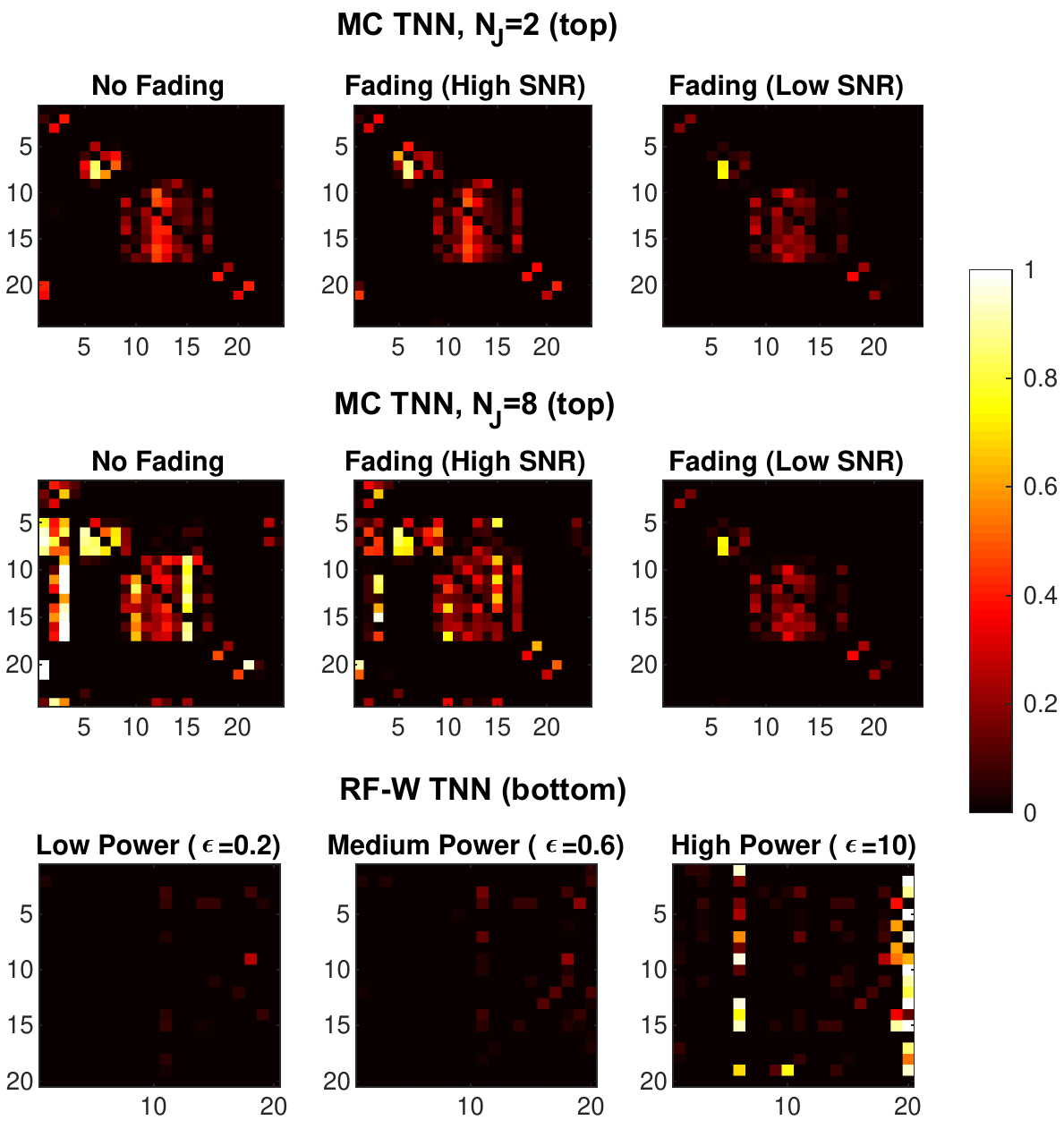}
    \caption{(top) Fooling matrix of MC TNN under AWJ-T for $N_J$ and $\epsilon$ values; (bottom) Fooling matrix of RF-W TNN under AWJ-T for different $\epsilon$ values and no fading.\vspace{-0.4cm}}
    \label{fig:jam_single_mod_matrix}
\end{figure}

Compared to untargeted jamming, AWJ-T requires smarter attack strategies as the adversary needs to (i) jam an already transmitted waveform, (ii) hide the underlying features of the jammed waveform and (iii) mimic those of another class. The top portion of Figure \ref{fig:jam_single_mod_matrix}  show the fooling matrices of AWJ-T attacks against MC TNN. Notice that \textbf{the higher the fooling rate, the more successful the attack is}. We notice that the adversary is able to effectively target a large set of modulations from 1 to 17 and 24 (\textit{i.e.}, OOK, M-QAM, M-PSK, ASK). However classes from 18-23 (\textit{i.e.}, AM, FM and GMSK) are hard to be targeted and show low fooling rate values. The bottom portion of Figure \ref{fig:jam_single_mod_matrix} shows  the results concerning the AWJ-T attack against RF-W TNN. In this case, the adversary achieves higher fooling rates with higher energy. 

\subsection{AML Waveform Synthesis (AWS) Results}\label{sec:aws_results}

Let us now evaluate the performance of AWS attacks in the case of rogue nodes. In this case, the attacker strategy $\boldsymbol{\phi}$  consists of $M$ complex-valued FIR taps (Section \ref{sec:synt}) that are convoluted with a baseband waveform $\mathbf{x}_{BB}$. To simulate a rogue device, we extract $\mathbf{x}_{BB}$ from the optimization set of the rogue class. This way we can effectively emulate a rogue class that needs to hide its own features and imitate those of the target classes. 

\begin{figure}[!h]
    \centering
    \includegraphics[width=0.9\columnwidth]{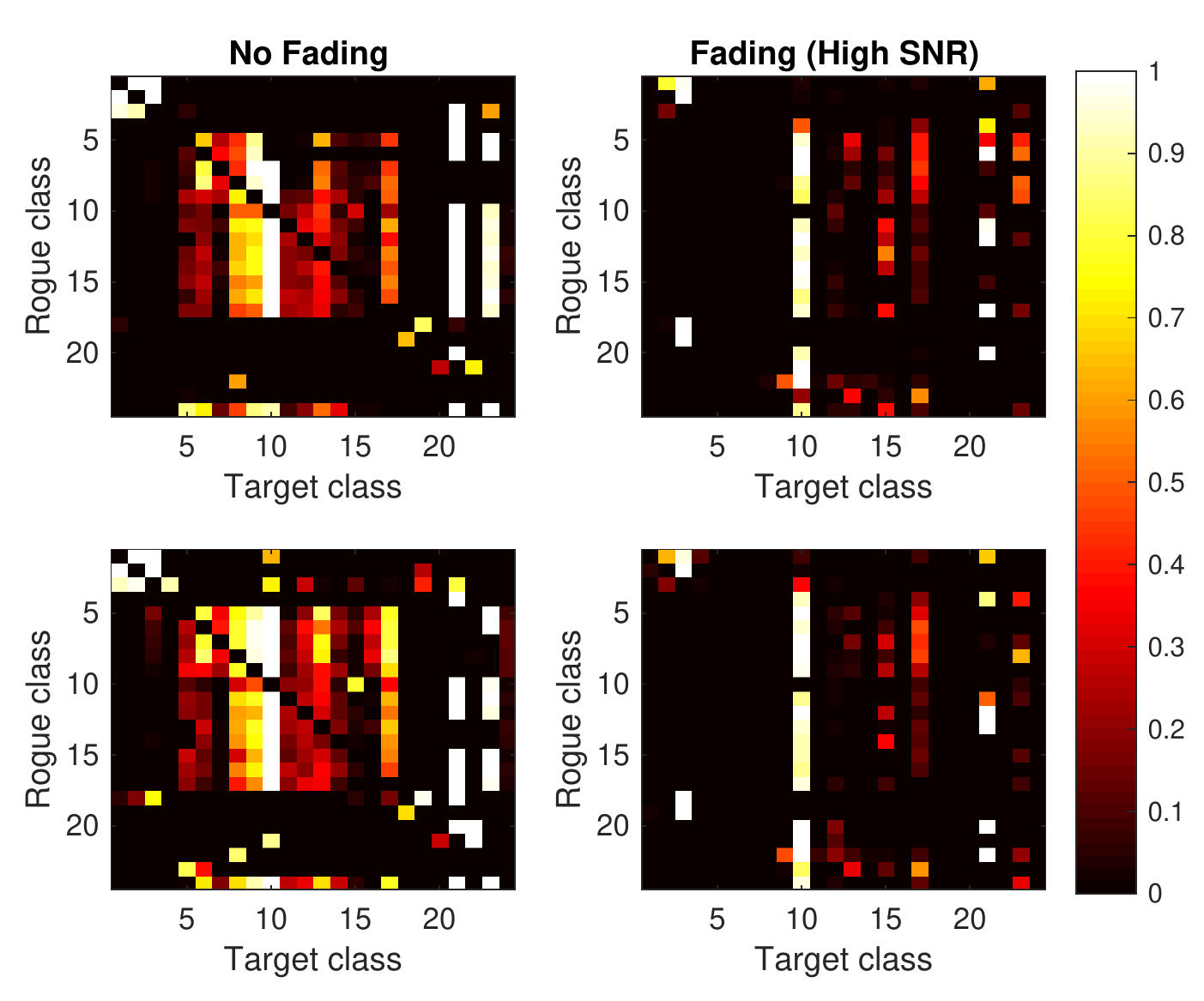}
    \caption{Fooling matrix of MC TNN under AWS with different $M$ ($M=4$: top; $M=8$: bottom).\vspace{-0.7cm}}
    \label{fig:synth_mod_matrix}
\end{figure}



\begin{figure}[!t]
    \centering
    \includegraphics[width=0.9\columnwidth]{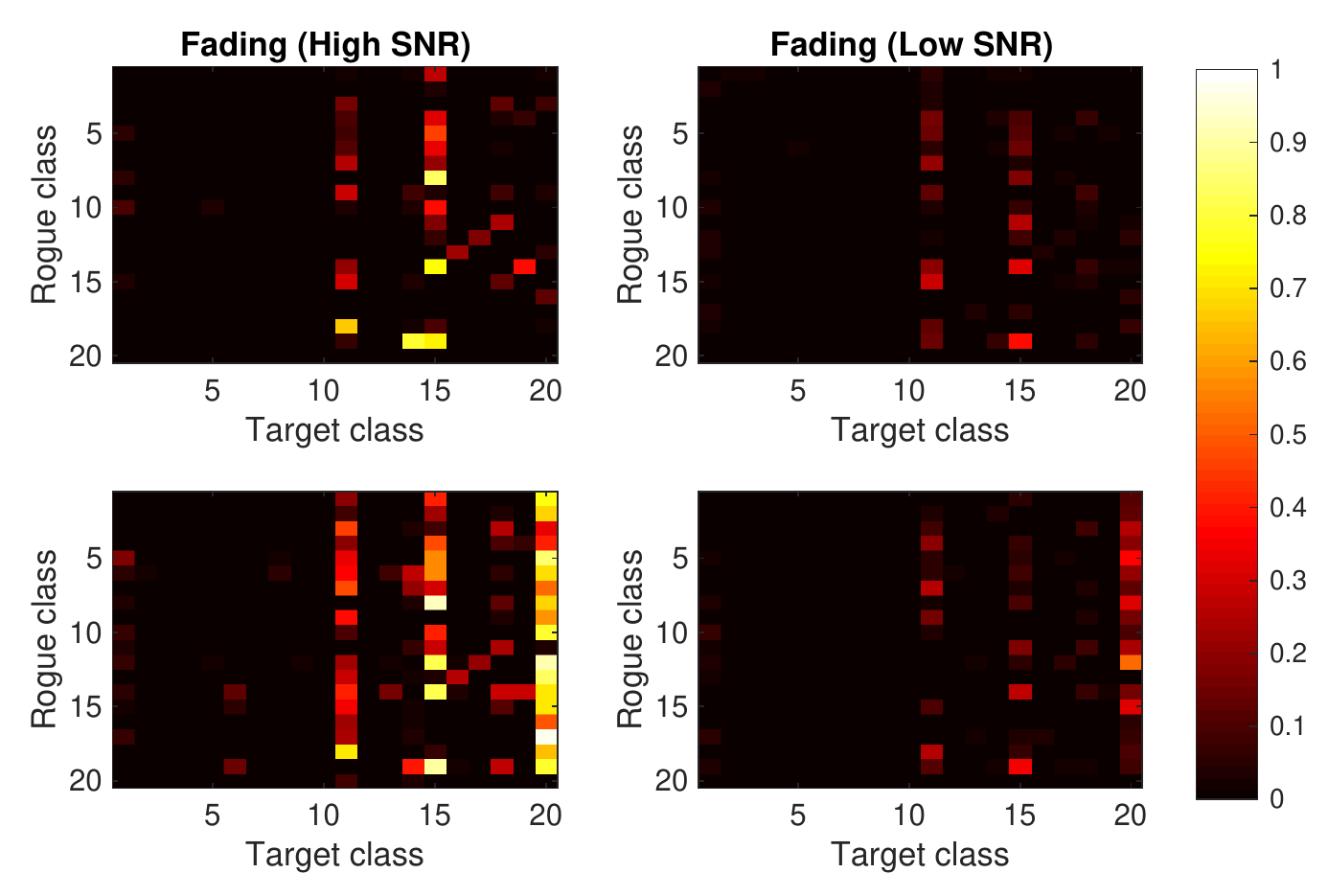}
    \caption{Fooling matrix of RF-W TNN under AWS for different values of $M$ ($M=4$: top; $M=8$: bottom).\vspace{-0.5cm}}
    \label{fig:synth_wifi}
\end{figure}

Figure \ref{fig:synth_mod_matrix} shows the fooling matrix of AWS attacks against the MC TNN for different channel conditions and values of $M$ when $\epsilon=0.2$. First, note that the main diagonal shows close-to-zero accuracy, meaning that the attacker can successfully hide its own features. Second, in the no-fading regime, rogue classes can effectively imitate a large set of target classes. Figure \ref{fig:synth_wifi} depicts the fooling matrices of AWS attacks against the RF-W TNN. We notice that (i) increasing the number of FIR taps increases the fooling rate; and (ii) \textbf{the bottom classes (1-10) are the ones that the attacker is not able to imitate}. However, the same does not hold for the top 10 classes (11 to 20), which can be imitated  with high probability (\textit{i.e.}, $28\%, 35\%, 62\%$ for classes 11,15,20, respectively). \textbf{Figure \ref{fig:synth_wifi} gives us an interesting insight on AWS attacks as it shows that the attacker is unlikely to attack those classes that are misclassified by the TNN.}

\begin{figure}[!h]
    \centering
    \includegraphics[width=0.9\columnwidth]{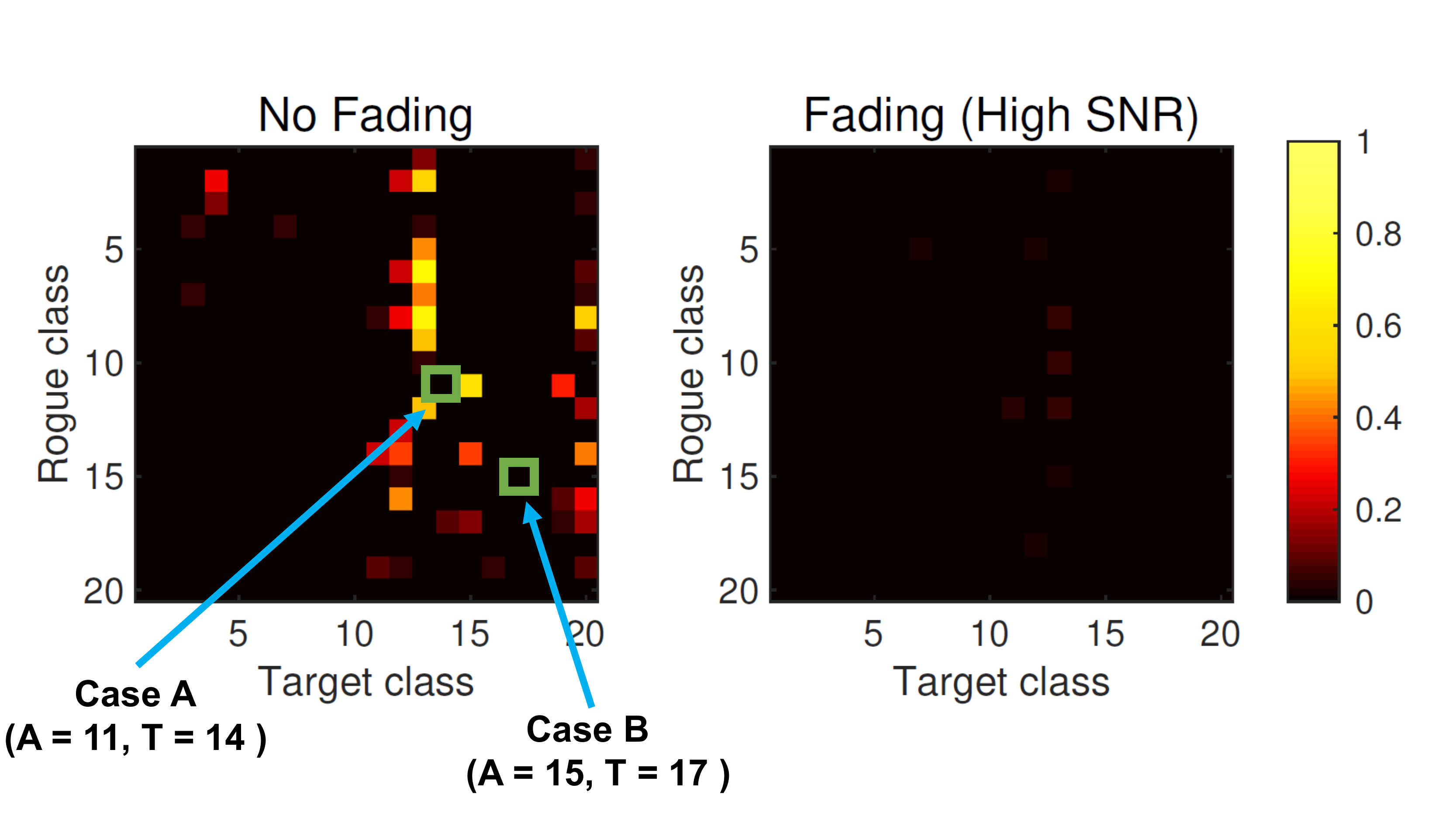}
    \caption{Fooling matrix of RF-A (original accuracy 60\%) TNN under AWS with $M=4$ and $\epsilon = 0.5$.\vspace{-0.3cm}}
    \label{fig:synth_adsb}
\end{figure}

The same behavior is also exhibited by the RF-A TNN. Figure \ref{fig:synth_adsb} shows the fooling matrix when $\epsilon = 0.5$ and $M=4$. Our results clearly show that the attacker is not able to properly imitate classes 1-10 (\textit{i.e.}, the bottom classes). Classes 11-20 (\textit{i.e.}, the top classes) can instead be imitated to some extent. This is because \textbf{it is unlikely that a unique setup of $\epsilon$ and $M$ will work for all classes} (both rogue and target). 

\begin{figure}[!h]
    \centering
    \includegraphics[width=0.9\columnwidth]{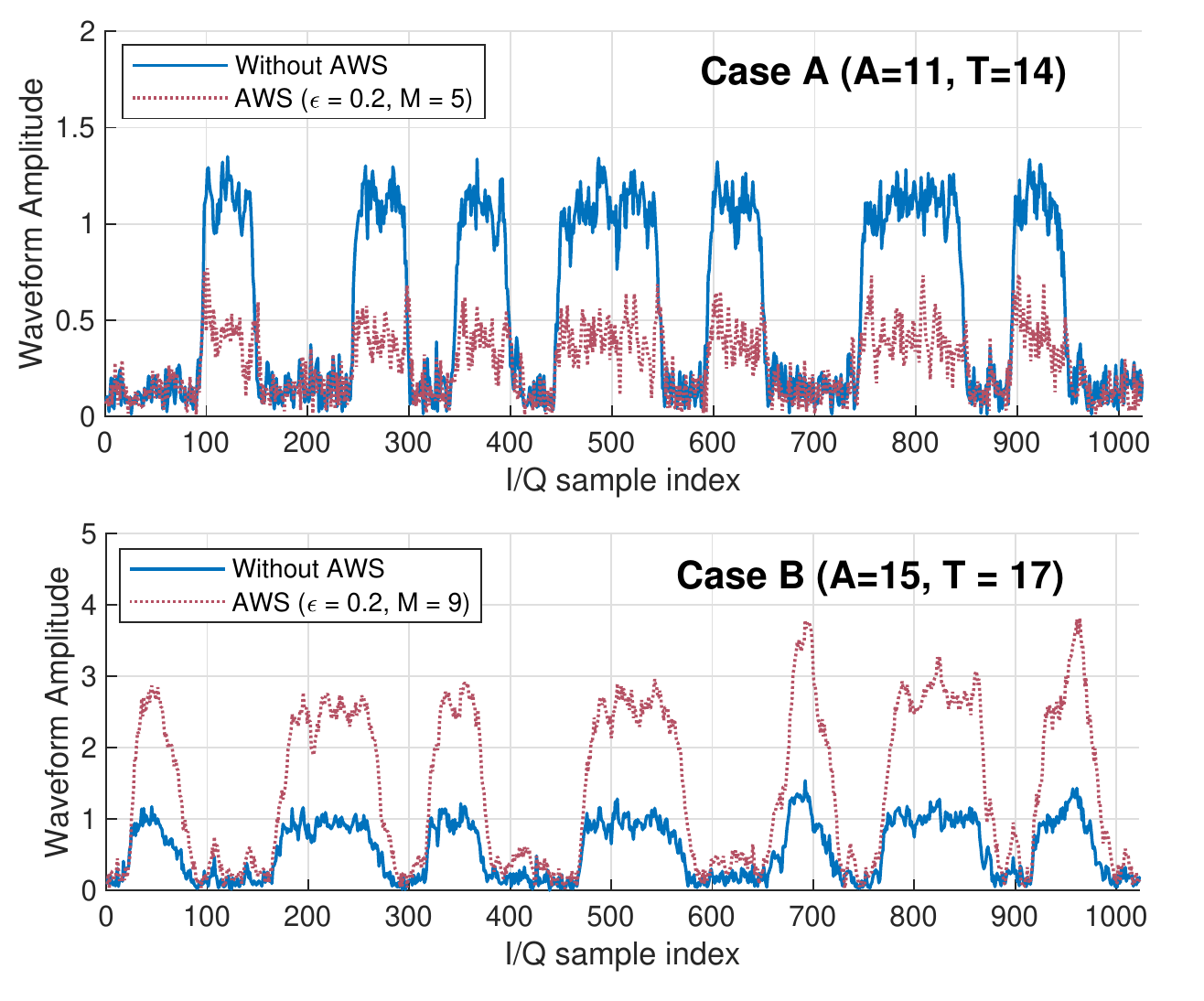}
    \caption{Comparison of waveforms generated through AWS attacks to RF-A TNN.\vspace{-0.8cm}}
    \label{fig:waveform}
\end{figure}

To further demonstrate this critical point, Figure \ref{fig:waveform} shows how rogue classes can actually imitate other classes by utilizing different values of $M$ and $\epsilon$. We define two cases: \textit{Case A}, where A=11 and T=14, and \textit{Case B}, where A=15 and T=17. As shown in Figure \ref{fig:synth_adsb}, Case A and B both yield low fooling rate when $M=4$ and $\epsilon = 0.5$. Figure \ref{fig:waveform} shows two ADS-B waveforms generated through AWS attacks in Case A and Case B, where solid lines show the original waveform transmitted by the rogue node without any modification in Case A and B. At first, the unmodified blue waveforms are classified by the RF-A TNN as belonging to the rogue class (11 and 15, respectively) with probabilities $97\%$ and $88\%$. However, by applying AWS with different M and $\epsilon$ parameters than the ones in Figure \ref{fig:synth_adsb}, the adversary is successful in imitating the target class in both Case A and B by increasing the activation probability to $20\%$ and $28\%$, \textit{which are considerably larger than the activation probability of all other 500 classes in the dataset}. This demonstrates that M and $\epsilon$ is critical to the success of the AWS.

Finally, the waveforms in Figure \ref{fig:waveform} give precious insights on how AWS actually operates. Interestingly, we notice that the phase of the waveforms does not change significantly, conversely from the amplitude. Since ADS-B uses an on-off keying (OOK) modulation, we verified that the modifications made by the waveform did not increase the BER of those transmissions. Moreover, Figure \ref{fig:waveform} shows that AWS attempts to change the path loss between A and R, as the amplitude respectively increases and decreases in Case A and B.

\subsection{FIRNet Testbed Results}\label{sec:over_the_air}

We evaluated \textit{FIRNet} on a software-defined radio (SDR) testbed composed by 64 omni-directional antennas through $100\:\mathrm{ft}$ coaxial cables. Antennas are hung off the ceiling of a $2240\:\mathrm{ft^2}$ office space and operate in the 2.4-2.5 and 4.9-5.9~GHz frequency bands. We pledge to share the collected waveform data and trained models with the community upon paper acceptance.


To evaluate the performance of \textit{FIRNet} in a challenging blackbox scenario, we implemented the targeted external Adversarial Waveform Synthesis (AWS) attack to a neural network used to fingerprint 5 nominally-identical USRP N210 radios transmitting an identical WiFi baseband signal. This is the \textit{worst-case} scenario for an adversary since \textit{FIRNet} has to learn the impairments to fool the neural network. The receiver SDR samples the incoming signals at $20\ \textrm{MS/s}$ and equalizes it using WiFi pilots and training sequences. The resulting data is used to train a TNN (see Figure 7 of \cite{restuccia2019deepradioid}) which takes as input 6 equalized OFDM symbols, thus 48*6 = 288 I/Q samples. It is composed by two 1D Conv/ReLU with dropout rate of 0.5 and 50 filters of size 1x7 and 2x7, respectively. The output is then fed to two dense layers of 256, and 80 neurons, respectively. We trained our network using the procedure in Section \ref{sec:training_models}. The resulting confusion matrix of the classifier, which obtains 59\% accuracy, is shown in Figure \ref{fig:conf_mat_intro}(a).

\begin{figure}[!h]
    \centering
    \includegraphics[width=0.9\columnwidth]{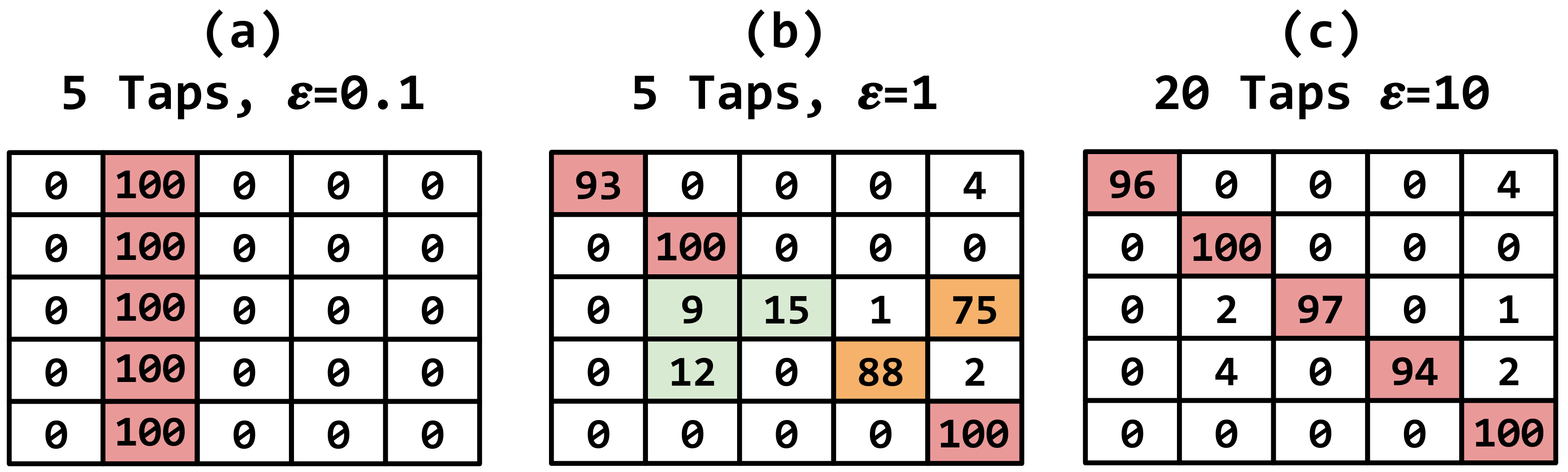}
    \caption{\emph{FIRNet} fooling matrices with 1 FIRLayer and different number of taps and $\epsilon$ value.\vspace{-0.6cm}}
    \label{fig:FIRNet_opt}
\end{figure}

We trained \textit{FIRNet} using baseband WiFi I/Q samples, thus without any impairment, with 1 FIRLayer and with a batch of 100 slices. Figure \ref{fig:FIRNet_opt}(a) shows that when $\epsilon$ has a low value of 0.1,  \emph{FIRNet}-generated I/Q sequences always collapse onto a single class, therefore are not able to hack the TNNs. However, Figure \ref{fig:FIRNet_opt}(b) shows that when $\epsilon$ increases to 1 the fooling rate jumps to 79\%, which further increases to 97\% with 20 FIR taps and $\epsilon = 10$, improving by over 60\% with respect to the replay attack that could achieve only 30\% fooling rate (see Figure \ref{fig:conf_mat_intro}(c)).

\begin{figure}[!h]
    \centering
    \includegraphics[width=0.9\columnwidth]{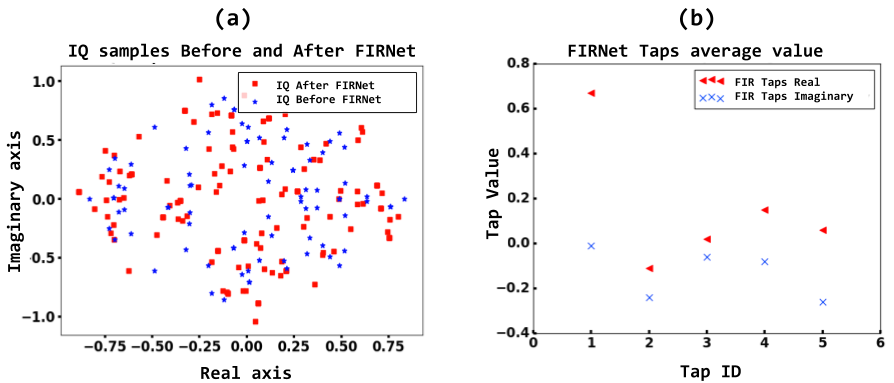}
    \caption{I/Q displacement and average FIR taps.\vspace{-0.2cm}}
    \label{fig:displacement}
\end{figure}

Finally, Figure \ref{fig:displacement}(a) and (b) show respectively the displacement caused by \textit{FIRNet} on an input slice with $\epsilon = 10$ and the average values of the 5 FIR taps obtained after training. We do not plot the remaining 15 taps since they are very close to zero. We notice that the distortion imposed to the I/Q samples is kept to a minimum, which is confirmed by the average FIR tap value which remains always below one.

\section{Conclusions} 

In this paper, we have provided a comprehensive, general-purpose modeling, analysis and experimental evaluation of wireless adversarial deep learning. First, we have formulate a Generalized Wireless Adversarial Machine Learning Problem (GWAP) to address AML in the wireless domain. Then, we have proposed algorithms to solve the GWAP in whitebox and blackbox settings. Finally, we have extensively evaluated the performance of our algorithms on existing state-of-the-art neural networks and datasets. Results demonstrate that our algorithms are effective in confusing the classifiers to a significant extent.



\small
\bibliographystyle{ieeetr}
\bibliography{main}

\end{document}